\documentclass{aa}

\usepackage{graphicx}
\usepackage{txfonts}
\newcommand{\onvire}[1]{}

\newcommand{\beq}{\begin{equation}}
\newcommand{\eeq}{\end{equation}}

\usepackage{color}
\usepackage{epsfig}
\usepackage{amssymb}
\usepackage{graphicx,natbib}
\usepackage{pdflscape}
\usepackage{longtable}
\usepackage[figuresright]{rotating} 
\begin{document}

\title{Sculpting the disk around T\,Cha: an interferometric view\thanks{Based on \textsc{Pionier} observations collected at the VLTI
    (European Southern Observatory, Paranal, Chile) with programs 087.C-0702(B), 087.C-0709(A), 089.C-0537(A), 083.C-0883(C \& D), and 083.C- 0295(A \& B).}}

   \author{J. Olofsson
          \inst{1}
          \and
          M. Benisty
          \inst{2}
          \and
          J.-B. Le Bouquin
          \inst{2}
          \and
          J.-P. Berger
          \inst{3}
          \and
          S. Lacour
          \inst{4}
          \and
          F. M\'enard
          \inst{5,2}
          \and
          Th. Henning
          \inst{1}
          \and
          A. Crida
          \inst{6}
          \and
          L. Burtscher
          \inst{7}
          \and
          G. Meeus
          \inst{8}
          \and
          T. Ratzka
          \inst{9}
          \and
          C.Pinte
          \inst{2}
          \and
          J.-C. Augereau
          \inst{2}
          \and
          F. Malbet
          \inst{2}
          \and
          B. Lazareff
          \inst{2}
          \and
          W. Traub
          \inst{10}
          }

   \offprints{olofsson@mpia.de}
   
   \institute{Max Planck Institut f\"ur Astronomie,
     K\"onigstuhl 17, 69117 Heidelberg, Germany \\
     \email{olofsson@mpia.de}
   \and
   UJF-Grenoble 1/CNRS-INSU, Institut de Plan\'etologie et d'Astrophysique de Grenoble (IPAG) UMR 5274, Grenoble, France
   \and
   European Southern Observatory, Alonso de Cordova, 3107, Vitacura, Chile
   \and 
   LESIA-Observatoire de Paris, CNRS, UPMC Univ. Paris 06, Univ. Paris-Diderot, 92195, Meudon, France
   \and
   UMI-FCA, CNRS / INSU France (UMI 3386) , and Departamento de Astronom\'ia, Universidad de Chile, Santiago, Chile.
   \and
   Universit\'e de Nice - Sophia Antipolis/C.N.R.S./Observatoire de la C\^ote d'Azur, Laboratoire Lagrange (UMR 7293), Boulevard de l'Observatoire, B.P. 4229 06304 NICE cedex 04, France
   \and
   Max-Planck-Institut f\"ur extraterrestrische Physik, Postfach 1312, Giessenbachstr., 85741 Garching, Germany
   \and
   Universidad Autonoma de Madrid, Dpt. Fisica Teorica, Campus Cantoblanco, Spain
   \and
   Universit\"ats-Sternwarte M\"unchen, Ludwig-Maximilians-Universit\"at, Scheinerstr. 1, 81679 M\"unchen, Germany
   \and
   Jet Propulsion Laboratory (NASA/JPL), MS 301-355, 4800 Oak Grove Drive, Pasadena, CA, 91109, USA
   }

   \date{Received \today; accepted }
 
   \abstract
   {Circumstellar disks are believed to be the birthplace of planets and are expected to dissipate on a timescale of a few Myr. The processes responsible for the removal of the dust and gas will strongly modify the radial distribution of the circumstellar matter and consequently the spectral energy distribution. In particular, a young planet will open a gap, resulting in an inner disk dominating the near-IR emission and an outer disk emitting mostly in the far-IR.}
   {We analyze a full set of data involving new near-infrared data obtained with the 4-telescope combiner (VLTI/\textsc{Pionier}), new mid-infrared interferometric VLTI/\textsc{Midi} data, literature photometric and archival data from VLT/NaCo/\textsc{Sam} to constrain the structure of the transition disk around T\,Cha.}
   {After a preliminary analysis with a simple geometric model, we used the \textsc{Mcfost} radiative transfer code to simultaneously model the SED and the interferometric observables from raytraced images in the $H$-, $L'$-, and $N$-bands.}
   {We find that the dust responsible for the strong emission in excess in the near-IR must have a narrow temperature distribution with a maximum close to the silicate sublimation temperature. This translates into a narrow inner dusty disk (0.07--0.11\,AU), with a significant height ($H/r \sim 0.2$) to increase the geometric surface illuminated by the central star. We find that the outer disk starts at about 12\,AU and is partially resolved by the \textsc{Pionier}, \textsc{Sam}, and \textsc{Midi} instruments. We discuss the possibility of a self-shadowed inner disk, which can extend to distances of several AU. Finally, we show that the \textsc{Sam} closure phases, interpreted as the signature of a candidate companion, may actually trace the asymmetry generated by forward scattering by dust grains in the upper layers of the outer disk. These observations help constrain the inclination and position angle of the disk to about $+58\degr$ and $-70\degr$, respectively.}
   {The circumstellar environment of T\,Cha appears to be best described by two disks spatially separated by a large gap. The presence of matter (dust or gas) inside the gap is, however, difficult to assess with present-day observations. Our model suggests the outer disk contaminates the interferometric signature of any potential companion that could be responsible for the gap opening, and such a companion still has to be unambiguously detected. We stress the difficulty to observe point sources in bright massive disks, and the consequent need to account for disk asymmetries (e.g. anisotropic scattering) in model-dependent search for companions.}
\authorrunning{Olofsson et al.}
\titlerunning{Sculpting the disk around T\,Cha}

   \maketitle
%

\section{Introduction\label{sec:intro}}

Circumstellar disks of young stellar objects are the products of in-falling matter during the process of star formation, and are believed to be the birthplace of planetary systems. These massive gas-rich, dusty disks will dissipate after a few Myr (\citealp{Hernandez2007}) and will then evolve into gas-poor debris disks, which contain the leftovers of planetary formation. As explained in the recent review by \citet{Williams2011}, disk evolution is driven by several mechanisms, from viscous transport to dust growth and sedimentation towards the midplane. These processes will contribute to the dissipation of the circumstellar material. In past decades, thanks to mid-infrared (mid-IR) spectroscopic campaigns (ISO and Spitzer, for instance), several disks experiencing a transition phase have been discovered (e.g., \citealp{Strom1989}; \citealp{Brown2007}; \citealp{Verhoeff2011}). The main characteristic of these transition disks is that they display a lack of emission in the mid-IR, while they show a strong excess in the far-IR. This has been interpreted as a sign of dust clearing in the inner regions, close to the star. Several processes can be responsible for the dissipation of the innermost material, such as the gravitational influence of a stellar companion (e.g., CoKu\,Tau\,4, \citealp{Ireland2008}), which can truncate the circumbinary disk. On a timescale of about $10^5$\,yr, the disk can also be dissipated from inside out by photo-evaporation (\citealp{Alexander2006}). The resulting density distribution will produce a spectral energy distribution (SED) similar to the one observed for the transition disk, where emission in excess becomes significant only at long wavelengths. A third possible scenario involves the presence of a recently formed planet in the disk, which will open a gap both in the gas and dust density distributions by pushing away surrounding material (e.g., \citealp{Papaloizou1984}). The mass of the forming planet will influence the size and depth of the gap (\citealp{Crida2006}), and the morphology of the outer disk (\citealp{Ayliffe2012}; \citealp{Pinilla2012}).

\object{T\,Cha} is one of the four transitional disks studied by \citet{Brown2007}, and it has a characteristic SED which displays a significant lack of emission in the mid-IR. \citet{Brown2007} modeled the circumstellar disk by defining two spatially separated regions, leaving a gap between 0.2 and 15\,AU. This gap was required to account for the lack of mid-IR emission. In \cite{Olofsson2011} we presented and modeled VLTI/\textsc{Amber} observations, with which we spatially resolved the inner dusty disk. We found this disk to be extremely narrow and close to the star (0.13--0.17\,AU). Given the small field of view of the interferometer (for this dataset, $\sim$\,60\,milli-arcsec, mas hereafter) and the overall low signal-to-noise ratio (S/N hereafter), the \textsc{Amber} data did not provide additional constraints on the outer disk. In \cite{Cieza2011}, we used Herschel/\textsc{Pacs} and \textsc{Spire} observations at 70, 160, 250, 350, and 500\,$\mu$m to study the structure of the outer disk (unresolved by the instruments). Because of the steep decrease in flux at far-IR wavelengths, we concluded the outer disk is fairly narrow with an outer radius between $\sim$20--30\,AU, depending on the surface density profile (between $-0.75$ and $-1$, respectively). 

Although the exact geometry of the system is revealed little by little as new observations are performed, they confirm there is a significant gap in the density distribution of the disk. This structure could be explained by the candidate companion detected by \cite{Huelamo2011}, which lies at a projected distance of 6.7\,AU from the star. As the companion pushes away surrounding material, it opens a large gap in the circumstellar disk, removing the dust content in the intermediate regions of the disk. However, models of gap opening by a planetary companion hardly explain the large size of the gap that observations suggest in the case of T\,Cha (\citealp{Zhu2012}). Thanks to the \textsc{Pionier} visitor VLTI instrument (\citealp{LeBouquin2011}), we revisit the disk structure.

In this study, we present new $H$- and $N$-band interferometric observations of T\,Cha (spectral type G8, \citealp{Alcal'a1993}) obtained with the \textsc{Pionier} and the \textsc{Midi} instruments (\citealp{Leinert2003}), both installed on the VLTI (\citealp{Haguenauer2010}). Additionally, we include archival VLT/NaCo/\textsc{Sam} observations (Sparse Aperture Masking, \citealp{Tuthill2010}) in our analysis. In the following, we first describe in section\,\ref{sec:obs} the observations and data processing and analyze the new interferometric data with a simple geometrical model (Sect.\,\ref{sec:geom}). In Sect.\,\ref{sec:rad}, we summarize the radiative transfer code capabilities and discuss how interferometric observables are derived from the code outputs. We then present and discuss our best fit model to the inner and outer disks (Sect.\,\ref{sec:inner} and \ref{sec:outer}, respectively), as well as the possible presence of a self-shadowed disk (Sect.\ref{sec:shadow}). We conclude in Sect.\,\ref{sec:discuss}.

\section{Observations and data processing\label{sec:obs}}

\begin{table}
\caption{Observations log and field-of-view ($\mathrm{FoV}$) for the different instruments.\label{tab:obs}}
\begin{center}
\begin{tabular}{lccc}
\hline \hline
Date & Baseline & Calibrator & Spectral mode \\
\hline
\multicolumn{4}{c}{VLTI/\textsc {Pionier}}\\
\multicolumn{4}{c}{$\mathrm{FoV} \sim 250$\,mas (25\,AU)}\\
\hline
2011-Apr-28 & D0-G1-H0-I1 & HD\,96494 & FREE \\
            &             & HD\,89615 & \\
2011-Jun-09 & A1-B2-C1-D0 & HIP\,57325 & SMALL-3 \\
& & HIP\,59243 &  \\
2012-Apr-15 &  A1-G1-I1-K0 & HIP\,57325 & FREE  \\
&  & HIP59243 & \\
\hline
\multicolumn{4}{c}{VLTI/\textsc{Midi}}\\
\multicolumn{4}{c}{$\mathrm{FoV} \sim 300$\,mas (30\,AU)}\\
\hline
2011-Mar-20 & UT2-UT3 & HD\,94717 & PRISM \\
2011-Mar-23 & UT2-UT3 & HD\,94717 & PRISM \\
2011-Mar-24 & UT1-UT4 & HD\,94717 & PRISM \\
\hline
\multicolumn{4}{c}{NaCo/\textsc{Sam}}\\
\multicolumn{4}{c}{$\mathrm{FoV} \sim 500$\,mas (50\,AU)}\\
\hline
2010-Mar-17 & L27 & HD\,102260 & $L'$ continuum\\
\hline
\multicolumn{4}{c}{VLTI/\textsc{Amber}}\\
\multicolumn{4}{c}{$\mathrm{FoV} \sim 60$\,mas (6\,AU)}\\
\hline
2009-Jun-11 & UT1-UT2-UT4 & HD\,107145 & LR \\
2009-Jun-12 & UT1-UT3-UT4 & HD\,107145 & LR \\
\hline
\end{tabular}
\end{center}
\end{table}%

\begin{figure*}
\begin{center}
\hspace*{-0.cm}\includegraphics[angle=0,width=2\columnwidth,origin=bl]{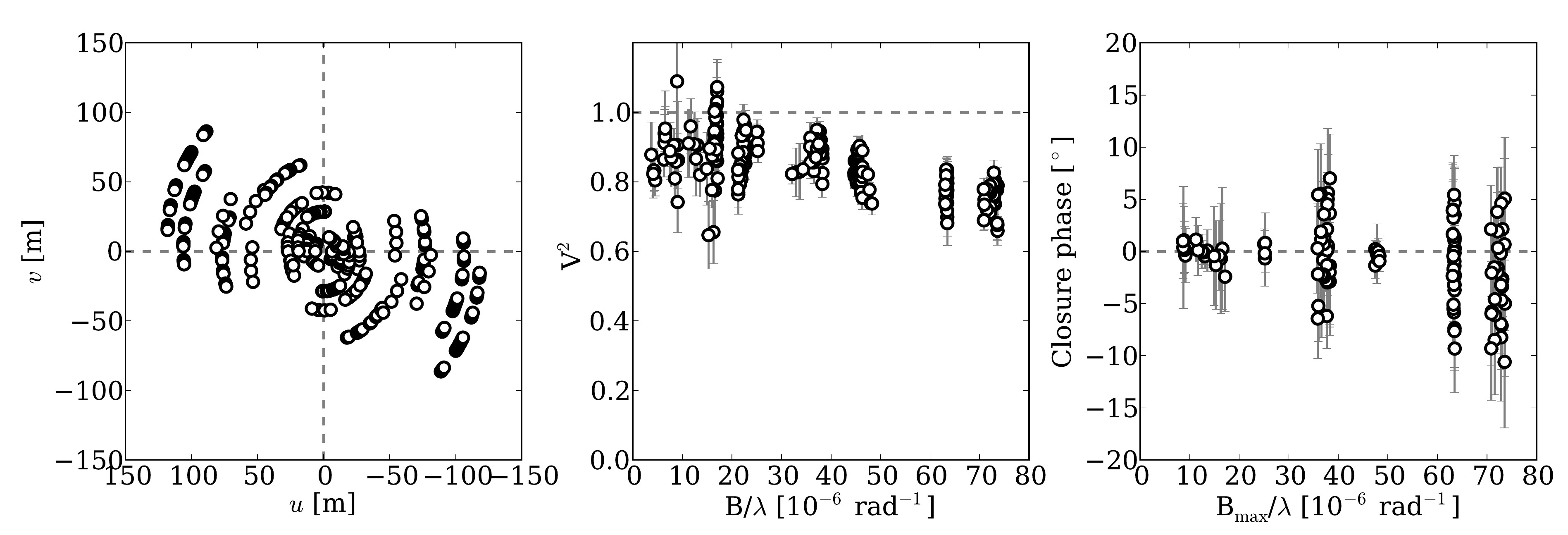}
\hspace*{-0.cm}\includegraphics[angle=0,width=\columnwidth,origin=bl]{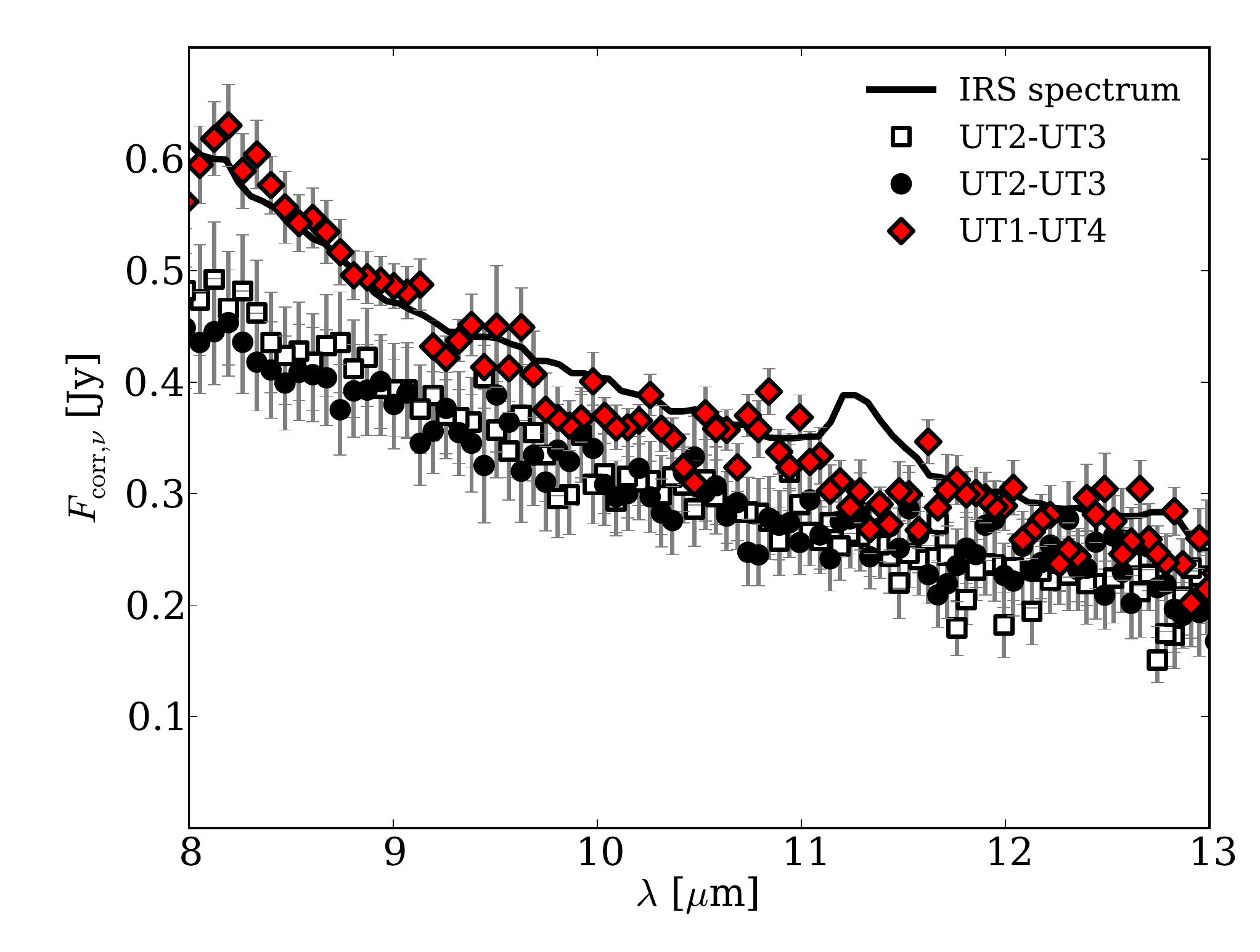}
\hspace*{-0.cm}\includegraphics[angle=0,width=\columnwidth,origin=bl]{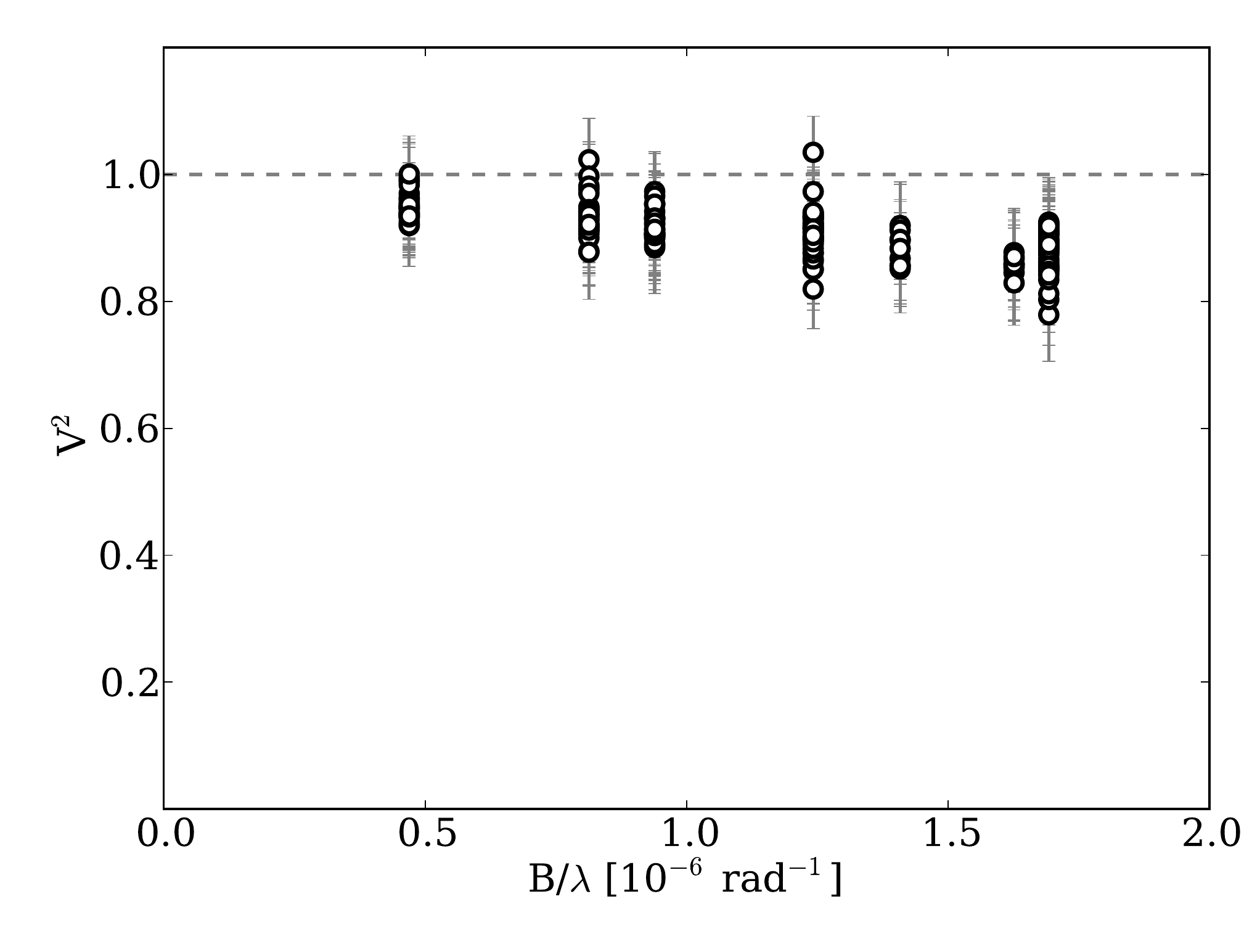}
\caption{\label{fig:data}{\it Top row:} from left to right: coverage of the $(u,v)$ plan, squared visibilities $V^{2}$ and closure phase for the \textsc{Pionier} observations. {\it Bottom left panel:} correlated fluxes for the three \textsc{Midi} observations and the \textsc{Irs} spectrum. Filled red diamonds (baseline UT1--UT4) have been discarded from the analysis (see text for details). {\it Bottom right panel:} NaCo/\textsc{Sam} squared visibilities as a function of spatial frequency.}
\end{center}
\end{figure*}

This study makes use of VLTI/\textsc{Pionier} observations (programs 087.C-0702(B), 087.C-0709(A), and 089.C-0537(A)), VLTI/\textsc{Midi} data (program 087.C-0883(C, D)), archival measurements obtained with NaCo/\textsc{Sam}, presented in \citet[][program 084.C-0755]{Huelamo2011}, and VLTI/\textsc{Amber} data already presented in \citet[][program 083.C- 0295]{Olofsson2011}. Table\,\ref{tab:obs} summarizes the observations and the instrumental setup for each instrument, as well as their field-of-view ($\mathrm{FoV}$ hereafter, which we assumed to be 2D Gaussians with different full width at half maximum, $\mathrm{FWHM}$).

\subsection{VLTI/\textsc{Pionier} observations}

Between April 2011 and April 2012, T\,Cha has been observed several times with the VLTI/\textsc{Pionier} instrument, with different configurations of the Auxiliary Telescopes (AT hereafter). The interferometer combines the light from four telescopes in the $H$-band, therefore simultaneously providing six squared visibilities ($V^2$ hereafter) and three independent (but four in total) closure phases per configuration. We aimed at a coverage of the $(u,v)$ plane as complete as possible, to probe different spatial scales of the circumstellar environment of T\,Cha. The minimum and maximum projected baselines are 6.3 and 123.8\,m (resolution of 55 and 2.8\,mas, respectively). Given the faintness of the star ($H = 7.9$\,mag), most of the observations were obtained without spectral dispersion (i.e stacking all the $H$-band light into a single detector pixel). The observing run of June 2011, however, was performed with the small spectral dispersion (3 channels across the $H$-band) and aimed at measuring precise closure phases to search for companions. Combined with the fact that the seeing conditions were below average, this explains the large uncertainties of these data points. Because of the large error bars, we spectrally binned the dispersed visibilities of this particular run for the rest of the analysis. 

Data reduction was performed using the standard pipeline ``\texttt{pndrs}" described in \citet{LeBouquin2011}. Figure\,\ref{fig:data} shows the calibrated data. From left to right, the upper three panels show the $(u,v)$ plane, $V^2$, and closure phases, respectively.

\subsection{VLTI/\textsc{Midi} observations}

The VLTI/\textsc{Midi} observations were taken in March 2011 using the Unit Telescopes (UT), in PRISM mode (spectral resolution of $R = 30$), and aimed at spatially resolving the distribution of polycyclic aromatic hydrocarbons (PAHs) in the circumstellar disk of T\,Cha. Data were processed using scripts dedicated to observations of faint targets with \textsc{Midi} (see \citealp{Burtscher2012b} for more details) interfaced with the \texttt{MIA+EWS} software\footnote{http://home.strw.leidenuniv.nl/$\sim$jaffe/ews/} (Version ``2012Jan25"). Bottom left panel of Figure\,\ref{fig:data} shows the calibrated correlated fluxes of T\,Cha as a function of wavelength for the three baselines and the Spitzer/\textsc{Irs} spectrum (aperture of 3.5$''$ for \textsc{Irs} versus $\sim$\,300\,mas for the \textsc{Midi} mask). Because T\,Cha is relatively faint for the instrument ($F_{\nu} < 1$\,Jy in the $N$-band), the \textsc{Midi} photometric observations were noisy and we prefer not to use them to compute visibilities from the correlated flux. One can note that both UT2-UT3 observations (black open squares and black filled circles, respectively), separated by 3 days, appear to be consistent with each other. As explained in \citet{Burtscher2012b} uncertainties are given by the quadratic sum of the photon noise error delivered by the \texttt{MIA+EWS} software and the standard deviation of the transfer function over the night of observations. Overall, the final uncertainties are dominated by the latter one (between 5 and 10\%). However, during the night of the 2011-Mar-24 (baseline UT1-UT4), only one \textsc{Midi} calibrated observation was performed, meaning we cannot estimate the standard deviation of the transfer function. Additionally, the fact that the correlated flux is higher for the longest baseline (UT1-UT4) is surprising, strongly suggesting an underestimation of the systematic errors. Because we cannot determine proper uncertainties for this dataset, we cannot assess if this deviation is real or not. 

\subsection{NaCo/\textsc{Sam} and VLTI/\textsc{Amber} literature data}

The \textsc{Sam} observations of T\,Cha were performed with VLT/NaCo on UT4, on the night of the 2010-Mar-17, with the L27 objective and a 7-hole mask. Data processing is described in \citet{Huelamo2011}, and the calibrated visibilities are shown in the bottom right panel of Figure\,\ref{fig:data}. The large uncertainties on the visibilities come from the longer exposure time to increase the signal in the originally prioritized closure phase signal. The candidate companion reported by \citet{Huelamo2011} was inferred from the $L'$-band closure phases. These are sensitive to the departure from centro-symmetry of the surface brightness distribution, and the authors found a minimum and a maximum of $-1.61$ and 1.69$^{\circ}$, respectively. The closure phase signature was attributed to the candidate companion by \citet{Huelamo2011}, and modeled as such. In the present paper the companion is not included. 

As discussed in \citet{Olofsson2011}, T\,Cha was observed with the VLTI/\textsc{Amber} instrument, in low-resolution mode (LR), during two consecutive nights of June 2009 (18 $V^2$ measurements on two different triplets, see Table\,\ref{tab:obs}). Details for data processing can be found in \citet{Olofsson2011}.

\subsection{Photometric literature data}

To construct the SED of T\,Cha, we used the photometric measurements, the Spitzer/\textsc{Irs} spectrum as in \citet{Olofsson2011}, and we included the Herschel/\textsc{Pacs} and \textsc{Spire} observations presented and modeled in \citet{Cieza2011}. The complete SED covers the spectral range between 0.36\,$\mu$m and 3.3\,mm.

\section{Preliminary analysis\label{sec:geom}}

Compared to the VLTI/\textsc{Amber} observations presented in \cite{Olofsson2011} the uncertainties of the \textsc{Pionier} $V^2$ in $H$-band are overall much smaller, however, the decrease in $V^2$ with increasing spatial frequencies is comparable. This drop in $V^2$ was interpreted and modeled as the signature of the inner dusty disk in \cite{Olofsson2011}. Interestingly, at lower spatial frequencies ($B/\lambda \leq$\,20\,$\times 10^{-6}$\,rad$^{-1}$) the $V^2$ display a ``plateau" ($V^2 < 1$) which suggests a contribution from an extended component, e.g. thermal emission from the disk that is later scattered or photospheric light scattered by the outer disk (see, \citealp{Pinte2008a} or \citealp{Tatulli2011} for a similar discussion applied to the case of HD\,100546).

\subsection{Ring model}

We first used a geometric model of an uniform ring (\citealp{Eisner2004}) on the VLTI/\textsc{Pionier} data for a preliminary analysis. The uniform ring model is described by three free parameters, its diameter $a$ (in mas), its position angle $PA$ ($0^{\circ}$ meaning the semi-major axis is oriented north-south), and its inclination $i$ ($0^{\circ}$ is face-on). The star-to-ring flux ratio is derived from the 2MASS $H$-band flux and the Kurucz stellar model obtained from SED fitting (see Sect.\,\ref{sec:rad}). We used a Monte Carlo Markov chain procedure (\texttt{emcee} package, \citealp{Foreman-Mackey2012}) to compute 120\,000 models. Probability density distributions for each of the three free parameters are then computed from the reduced $\chi_{\mathrm{r}}^2$ (projection of $\mathrm{exp}[-\chi^2_{\mathrm{r}}/2]$ along the three dimensions of the parameter space). The most probable fit to the data is obtained for a diameter $a = 1.9\pm0.1$\,mas (0.19$\pm0.01$\,AU), a position angle of $142^{\circ}\pm4$, and an inclination $i = 53^{\circ}\pm2$ (the uncertainties correspond to the width $\sigma$ of a Gaussian fit to the probability density distributions). The inferred inclination of 53$^{\circ}$ is close to our result in \citet{Olofsson2011}, and close to the most probable inclination found for the outer disk (\citealp[$i \sim$\,60$^{\circ}$,][]{Cieza2011}), suggesting that within the uncertainties, the inner and outer disks are coplanar.

\subsection{No detected companion in the $H$-band}

Additionally, we checked for a possible binary signal in the \textsc{Pionier} data. We used the procedure described in \citet{Absil2011}. We do not detect any signal for a companion, down to a flux ratio of 1:100 in the $H$-band, for separations in the range 10--100\,mas (1--10\,AU).

\section{Radiative transfer code and interferometric quantities\label{sec:rad}}

\subsection{The \textsc{Mcfost} code}

To have a more comprehensive understanding of the circumstellar environment of T\,Cha, we used the radiative transfer code \textsc{Mcfost} (\citealp{Pinte2006,Pinte2009}) to simultaneously reproduce the SED and the interferometric observables. Raytraced images are produced for the $H$-, $L'$-, and $N$-band and used to compute $V^2$, closure phases, and correlated fluxes. In \textsc{Mcfost} a disk is defined with the following parameters: the inner and outer radii ($r_{\mathrm{in}}$, $r_{\mathrm{out}}$, respectively), an exponent $\alpha$ for the surface density distribution ($\Sigma( r ) = \Sigma_0 (r/r_0)^{\alpha}$) and the mass of the dust $M_{\mathrm{dust}}$. The vertical structure follows a Gaussian distribution (exp$[-z^2 / 2H( r )^2]$), where the scale height $H(r)$ is defined with a reference value $H_0$ at the reference radius $r_0$. The flaring of the disk is described via the exponent $\beta$  ($H( r ) = H_0(r/r_0)^{\beta}$). Finally the dust grain properties are fully described by the optical constants (used to compute mass absorption coefficients for a given dust composition), the minimum and maximum grain sizes ($s_{\mathrm{min}}$ and $s_{\mathrm{max}}$, respectively), and a power-law exponent $p$ for the grain size distribution ($dn(s) \propto s^p$d$s$, $p < 0$). Several, spatially separated regions can be defined in \textsc{Mcfost}. The monochromatic raytraced images include the thermal and scattered light contributions.

To define the stellar photosphere within \textsc{Mcfost}, we used the same stellar parameters as in \citet[][$T_{\mathrm{eff}}$=\,5400\,K, $d$=100\,pc, $A_{\mathrm{v}}$=1.5]{Olofsson2011}. We assumed the geometry of the disk to be the same as the one originally described in \citet{Olofsson2011}, with an inner dusty disk close to the star that is responsible for the near-IR excess and an outer disk located further away from the star that contributes to the far-IR. 

The \textsc{Mcfost} model is hardly comparable to an infinitely flat disk as assumed in the case of the geometric model, therefore the  parameters found using this model ($i$, $a$ and the $PA$) for the inner disk may be different for the \textsc{Mcfost} model. Given that the emission is not fully resolved by \textsc{Pionier} ($V^2 \sim 0.7$), $i$ and $PA$ cannot be fully constrained by these measurements. However, as discussed later on (see Sect.\,\ref{sec:outer}), the outer disk is in the $\mathrm{FoV}$ of \textsc{Sam} and its inclination and position angle can be further constrained. In this study, we will assume the inner and outer disks to be coplanar.

\subsection{Computing visibilities and closure phases}

With \textsc{Pionier}, we observed T\,Cha using the ATs, which have a photometric field of view that can be approximated by a 2D Gaussian with a $\mathrm{FWHM}$ of about 250\,mas. Additionally, even in its broad-band mode \textsc{Pionier} is not a monochromatic instrument, but covers a spectral range within the $H$-band. One has to be careful when computing $V^2$ from raytraced images, especially when the photometric $\mathrm{FoV}$ is large. We can define the photometric flux as
\begin{eqnarray}\label{eqn:phot}
F_{\mathrm{phot}} = \int_{\theta} \int_{\lambda} I(\theta,\lambda) \, \mathrm{d}\theta \, \mathrm{d}\lambda,
\end{eqnarray}
and the coherent flux as
\begin{eqnarray}\label{eqn:coh}
F_{\mathrm{coh}} = \int_{\theta} \int_{\lambda} I(\theta,\lambda) e^{-2i\pi \theta B / \lambda} \, \mathrm{d}\theta \, \mathrm{d}\lambda,
\end{eqnarray}
where $I(\theta)$ is the source intensity as a function of the centro-symmetric coordinate on the sky $\theta$, $B$ and $\lambda$ are the projected baseline and wavelength of observations, respectively. Then the resulting visibility $V$ will be the ratio $F_{\mathrm{coh}}/F_{\mathrm{phot}}$. Under the reasonable assumption that the raytraced image is achromatic within the $H$-band, Eq.\,(\ref{eqn:phot}) becomes
\begin{eqnarray}\label{eqn:phot2}
F_{\mathrm{phot}} = \int_{\theta} I(\theta) \, \mathrm{d}\theta.
\end{eqnarray}
However, this simplification cannot be done in Eq.\,(\ref{eqn:coh}) because of the wavelength dependent term in the exponential. For small baselines $B$, the differences in spatial frequencies within the $H$-band will be small and the effect negligible. However, for longer baselines, this effect becomes significant, especially when the emission is resolved. To account for this wavelength dependency, we computed the complex visibilities at different wavelengths (21 points between $\lambda_{\mathrm{min}} = 1.28$ and $\lambda_{\mathrm{max}} = 2.08$\,$\mu$m) assuming the raytraced image at $\lambda_0 = 1.68$\,$\mu$m to be identical within this spectral range ($I(\theta,\lambda)$ can be approximated by $I(\theta)$). The complex visibilities were then averaged before computing the $V^2$.

Concerning \textsc{Sam} and \textsc{Midi} observations, we proceeded in a similar way to account for chromaticity of the measurements. We additionally made sure to adapt the size of the raytraced images to the $\mathrm{FoV}$ of the instruments (500 and 300\,mas, respectively). To compute \textsc{Sam} $V^2$ and closure phases, we took 21 wavelength points between 3.2 and 4.4\,$\mu$m. To compare our model to the \textsc{Midi} data, we computed the correlated fluxes at three wavelength points: 9, 10 and 12\,$\mu$m (observed data were averaged over a $\pm 0.5$\,$\mu$m range) in order to see if we could reproduce the wavelength dependency of the observations. Therefore, we generated raytraced images at each monochromatic wavelength, computed visibilities at the spatial frequency of the observations, over a spectral range of $\pm 0.5$\,$\mu$m as discussed in the previous paragraph, and multiplied the averaged visibilities by the total flux in the images. The spectral range of $\pm 0.5$\,$\mu$m was chosen to be the same with which we averaged the observed correlated fluxes. The complete analysis of the \textsc{Midi} data, including the PAHs feature, will be presented in Meeus et al. (in preparation).

\section{Modeling the inner disk\label{sec:inner}}

In \citet{Olofsson2011} we showed that the inner disk is responsible for both the decrease of the $V^2$ and the near-IR excess, therefore any attempt to model it has to be led carefully through a combined approach on the interferometric data and the SED, pending the acquisition of a direct image. In the following we underline some of the particular features of the inner disk that are inferred from the SED.

\subsection{The near-IR excess\label{sec:inner_excess}}

\begin{figure}
\begin{center}
\hspace*{-0.cm}\includegraphics[angle=0,width=1\columnwidth,origin=bl]{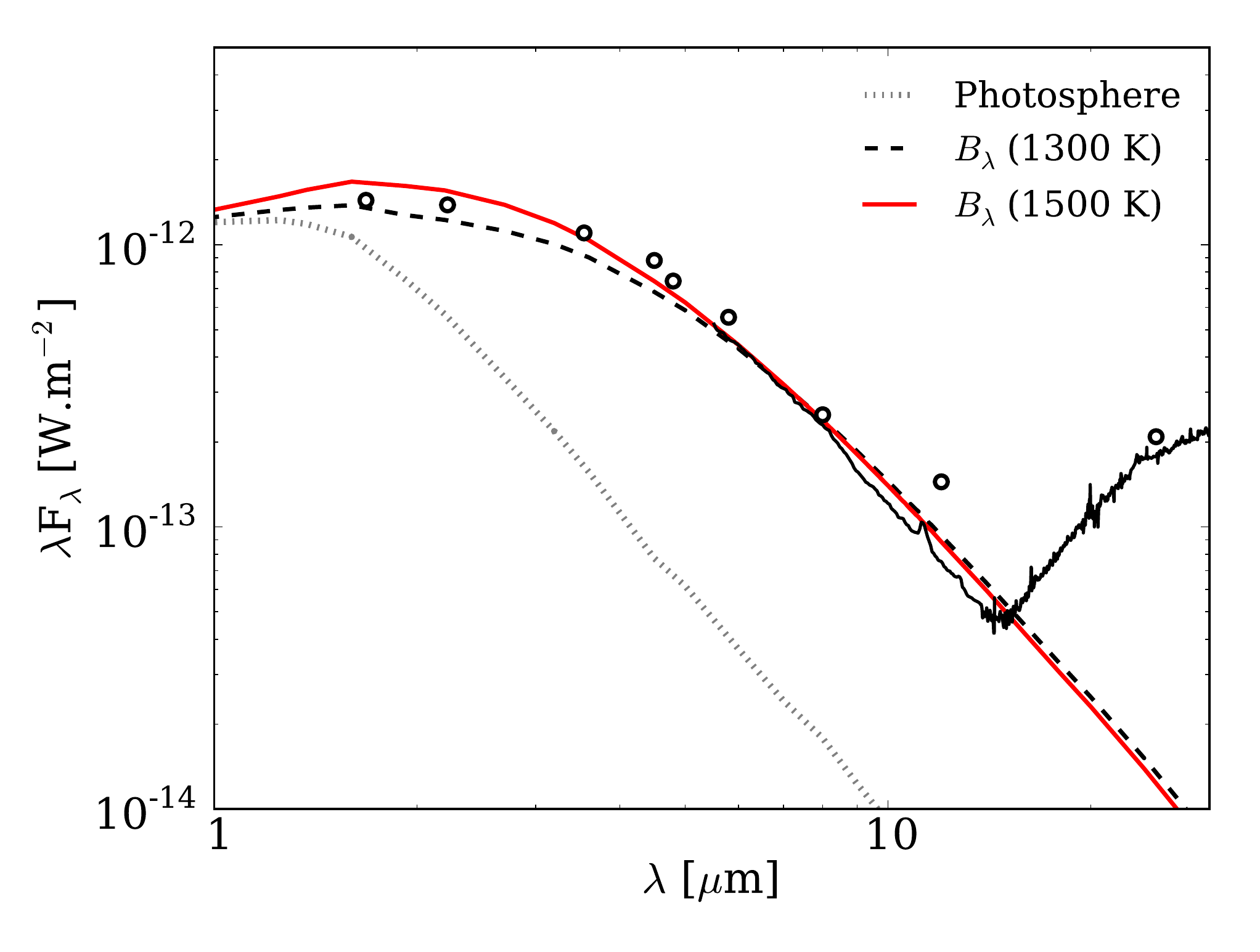}
\caption{\label{fig:inner}Blow-up of the SED of T\,Cha in the near-, mid-IR. Stellar photosphere is shown in dotted grey. Two Planck functions at different temperatures ($T = 1300$ and 1500\,K) are shown in dashed black and solid red, respectively. Both Planck functions are scaled to the \textsc{Irs} flux at 6\,$\mu$m.}
\end{center}
\end{figure}

The near-IR excess around T\,Cha is already significant in the $H$-band (about 25\% of the stellar light), suggesting a large amount of warm dust is in the immediate vicinity of the star. Figure\,\ref{fig:inner} shows a blow-up of the SED in the near- to mid-IR (1--30\,$\mu$m), with the stellar photosphere and the \textsc{Irs} spectrum. Additionally, to have a first estimate of the characteristic temperature of the dust responsible for the near-IR excess, two Planck functions at $T$\,=\,1300 and 1500\,K are shown to mimic thermal emission at high temperatures. Both Planck functions are added to the stellar photospheric model and are scaled to match the \textsc{Irs} flux at 6\,$\mu$m. One can see that the near-IR excess is slightly under- and over-estimated for temperatures of 1300, 1500\,K, respectively. Most importantly, both Planck functions slightly over-predict the \textsc{Irs} flux in the range 8--15\,$\mu$m, suggesting a decrease with wavelength steeper than a Planck function in the mid-IR regime. For this simple exercise, we only considered a single temperature for the dust grains, while they most likely have a distribution of temperatures. The inclusion of dust at temperatures colder than 1300\,K will increase the emission in excess at about 15\,$\mu$m, which underlines the complexity of reproducing the sharp ``V-shaped" dip in the mid-IR. Overall this suggests that a vast majority of the dust grains has a single, warm ($T > 1300$\,K) temperature. This can be explained if the inner disk becomes rapidly optically thick (as discussed in \citealp{Grady2009} in the case of HD\,135344\,B).

\subsection{Dust composition\label{sec:inner_comp}}

The \textsc{Irs} spectrum of T\,Cha does not show any emission features associated with silicate grains, especially at 10\,$\mu$m, where most spectra of circumstellar disks would show the emission band arising from the Si--O stretching vibrations in amorphous silicate dust grains (see \citealp{Henning2010} for a review of cosmic dust optical properties). There could be two different ways to suppress the 10\,$\mu$m emission feature: {\it (i)} the inner disk does not contain any silicate dust grains, {\it (ii)} silicate grains are typically larger than 5--10\,$\mu$m. Indeed, once silicate grains have grown to size larger than several $\mu$m, they no longer display the 10\,$\mu$m emission feature. To some extent, this effect can also be mimicked by high porosity or carbon inclusion in the silicate matrix (see \citealp{Voshchinnikov2008} for instance). 

First, scenario {\it (i)} can be ruled out fairly easily as it would be counter-intuitive not to have any silicate dust grains in the inner disk. Silicate grains have been observed in numerous astronomical environments: the winds of AGB stars, the interstellar medium, protoplanetary disks, and planetary crusts in the solar system. To match the near-IR excess of T\,Cha, dust temperatures do not necessarily need to be higher than the dust sublimation temperature ($T \sim 1500$\,K, see Fig.\,\ref{fig:inner}). Therefore, there is no immediate need to use only refractory dust grains (i.e., resistant to higher temperatures). One should note that we do not (and cannot) rule out their presence: there is simply no observational evidence that requires their inclusion in the model.

Consequently, we examined in detail the second possibility in which the inner disk contains silicate grains with sizes larger than 0.1\,$\mu$m. We used optical constants from \citet{Draine1984} to represent the emission properties of silicate dust grains, and did not include any other dust species. We fixed the maximum grain size $s_{\mathrm{max}}$ to 1000\,$\mu$m and varied $s_{\mathrm{min}}$ (always assuming $p = -3.5$ for the grain size distribution, \citealp{Dohnanyi1969}). For a given dust composition, the maximum temperature reached is governed by $s_{\mathrm{min}}$; the smaller the grains, the warmer. As discussed previously, to obtain a satisfying fit to the near-IR excess, dust grains need to be warm enough ($1300 < T <1500$\,K). Therefore, the position of the inner radius of the inner disk is, as a first approximation, driven by the minimum grain size. With $s_{\mathrm{min}} \leq 5$\,$\mu$m, the 10\,$\mu$m emission feature is always detected in the models. For $s_{\mathrm{min}} = 5$\,$\mu$m, the dust has to be located close to the star ($r_{\mathrm{in}} \sim 0.04$\,AU). This model results in a poor fit to the \textsc{Pionier} $V^2$, especially at long baselines where the modeled $V^2$ over-predict the observed squared visibilities. Indeed, as shown in Sect.\,\ref{sec:geom}, the typical diameter probed by the \textsc{Pionier} data is of about 2\,mas ($r_{\mathrm{in}}$=0.1\,AU at 100\,pc). In other words, dust grains must be located at about $\sim$0.1\,AU and have a temperature between 1300--1500\,K. This condition is required to simultaneously model the SED and the $V^2$ and can only be achieved if small grains ($s \lesssim 0.1$\,$\mu$m) are included in the model.

Similar investigations were led by \citet{Meeus2002} who modeled the SED of HD\,100453. The ISO spectrum for this source does not show any 10\,$\mu$m emission feature. The authors therefore modeled the spectrum of this source using amorphous grains with olivine stoichiometry and carbon grains, both species having sizes between 4 and 200\,$\mu$m. The authors also included metallic iron grains with sizes between 0.01 and 2.5\,$\mu$m. The justification for including this dust species is that dust temperatures higher than the sublimation temperatures of the silicates and carbon grains were mandatory to reproduce the near-IR excess. This is not the case for T\,Cha. We conducted similar tests, trying to address the following question: how much silicate can we hide in the inner disk before the 10\,$\mu$m feature becomes detectable? In addition to the optical constant of astro silicates from \citet{Draine1984}, we included amorphous carbon (\citealp{Jager1998}). We choose amorphous carbon, because its mass absorption coefficients are featureless at all sizes. Grain sizes for carbon were set to $0.01 < s < 1000$\,$\mu$m, and $s_{\mathrm{max}} = 1000$\,$\mu$m for amorphous silicates grains (with $p = -3.5$ for both dust species). We varied $s_{\mathrm{min}}$ for the astro silicates, as well as the relative mass fraction between the two dust species and aimed at reasonable fits to both the SED and \textsc{Pionier} $V^2$. We had to change the disk geometry in order to test the robustness of each dust setup. We find that a relative mass fraction of 70\% of astro silicates, with a minimum size of 5\,$\mu$m leads to a good fit to all the observables. With less carbon grains, $r_{\mathrm{in}}$ has to be too close to the star for the $V^2$ to be fitted properly. With smaller $s_{\mathrm{min}}$ for the silicate dust grains, a 10\,$\mu$m emission feature becomes detectable.

\subsection{Disk parameters\label{sec:inner_mod}}

\begin{table}
\caption{Range of parameters explored for the modeling of the inner and outer disks. See Sect.\,\ref{sec:inner_mod} and \ref{sec:disk_outer} for details.\label{tab:inner}}
\begin{center}
\begin{tabular}{lc}
\hline \hline
Parameter & Range \\
\hline
\multicolumn{2}{c}{Inner disk} \\
\hline
$r_{\mathrm{in}}$ & $[$0.06, 0.07, 0.08, 0.09, 0.1, 0.11$]$\,AU \\
$W = r_{\mathrm{out}} - r_{\mathrm{in}}$ & $[$0.03, 0.04, 0.05, 0.07, 0.1$]$\,AU \\
$H_0$ ($r_0=0.1$\,AU) & $[$0.015, 0.0175, 0.02, 0.0225$]$ \\
$M_{\mathrm{dust}}$ & $[$ 0.9, 1, 2, 3, 4, 7.5, 10$] \times 10^{-11}$\,$M_{\odot}$ \\
$i$ & [50, ..., 70]$\degr$, 1$\degr$ steps \\
$PA$ & [0, ..., 180]$\degr$, 10$\degr$ steps \\
\hline
\multicolumn{2}{c}{Outer disk} \\
\hline
$r_{\mathrm{in}}$ & $[$8, 9, 10, 11, 12, 13, 14, 15$]$\,AU \\
$H_0$ ($r_0 = 25$\,AU) & $[$1.5, ..., 2.5$]$\,AU, 0.1\,AU steps \\
$M_{\mathrm{dust}}$ & $[$1, 5, 6, 7, 8, 9, 10, 20$] \times 10^{-5}$\,$M_{\odot}$ \\
$i$ & [50, ..., 70]$\degr$, 1$\degr$ steps \\
$PA$ & [-180, ..., 180]$\degr$, 1$\degr$ steps \\
\hline
\end{tabular}
\end{center}
\end{table}

As discussed in Sect.\,\ref{sec:inner_excess}, to match the fast decrease of the flux between 2 and 15\,$\mu$m, dust grains must have a narrow range of temperatures, centered around 1300--1500\,K. Allowing for colder grains in the inner disk would increase the flux at about 15\,$\mu$m. The best fit model for the inner disk is found within a grid of models that aim at reproducing both the SED, for wavelengths between 0.3 and 15\,$\mu$m, and the \textsc{Pionier} visibilities, for which we used the entire dataset. For the SED, we used 13 photometric points (the fluxes at \textsc{Irs} wavelengths were interpolated at 10 and 13\,$\mu$m). We searched for the best models by minimizing the un-weighted sum of reduced $\chi^2_\mathrm{r}$ for the SED and \textsc{Pionier} $V^2$. The uncertainties quoted for each parameters are derived from the marginalized probability distributions ($\propto \mathrm{exp}[-\chi^2_\mathrm{r}/2]$). The search is performed on the following parameters: $r_\mathrm{in}$, $r_\mathrm{out}$, $H_0$, and $M_\mathrm{dust}$. For each model, raytraced images are computed for 21 inclinations, and the $V^2$ are fitted for 19 values of the $PA$. Table\,\ref{tab:inner} summarizes the range of parameters explored within the grid. As previous analysis (\citealp{Brown2007}; \citealp{Olofsson2011}) have shown that the inner disk is narrow, we fixed the surface density to $\alpha = -1$ and the flaring exponent $\beta = 1$ as they have little to no influence on the resulting SED or $V^2$. Given that the \textsc{Midi} data are used to check for consistency, they are not included in the search for the best fit model. Given the much better $(u,v)$ coverage and S/N of the \textsc{Pionier} data compared to the \textsc{Amber} data published in \citet{Olofsson2011}, the latter ones are not included in the $\chi^2_\mathrm{r}$ minimization process.

We find the inner disk to be best modeled for $r_{\mathrm{in}} = 0.07 \pm 0.01$\,AU and $r_{\mathrm{out}} = 0.11 \pm 0.02$\,AU, values consistent with the results from the geometric model and those from \citet{Olofsson2011}, where we used a slightly different dust composition and the data had a lower S/N. We could obtain comparable fits to the \textsc{Pionier} $V^2$ with a larger width for the inner disk ($r_{\mathrm{out}}$ up to 0.15--0.2\,AU), but the flux in the near- to mid-IR then becomes redder and slightly over-predicts the flux at 10--15\,$\mu$m, while the flux at shorter wavelengths is under-predicted. The scale height is a crucial parameter. To obtain a majority of dust grains at high temperatures, with very few grains at temperatures below $\sim$\,1000\,K, one can increase the scale height of the disk. A larger geometric cross-section of the inner disk increases the number of grains directly exposed to the stellar radiation, which are therefore efficiently heated up. At a reference radius $r_0$ of 0.1\,AU, we obtained a good match to the data by setting the scale height at $H_0 =\,0.02 \pm 0.0025$\,AU. For such a model, we obtained a maximum temperature of $T_{\mathrm{dust}} \sim$ 1510\,K, close to the sublimation temperature of sub-\,$\mu$m-sized silicate dust grains. A total dust mass of about $2_{-1}^{+2} \times 10^{-11}$\,$M_{\odot}$ is sufficient to match the SED in the near-IR. The best inclination and position angle, for the inner disk solely, are found to be $54^{+6}_{-4}$\,$\degr$, and $120\degr \pm 30\degr$, respectively. The latter two values will be revisited when modeling the outer disk (Sect.\ref{sec:outer}). The relatively large uncertainty on the position angle comes from the fact that the emission is not fully resolved by \textsc{Pionier} ($V^2 \geq 0.7$).
 
\subsection{Discussion}

T\,Cha is not the only transition disk showing a lack of silicate emission features in the mid-IR (e.g., HD\,135344\,B, SR\,21). In the case of T\,Cha, the maximum temperature reached in the best-fit model is slightly higher than 1500\,K. We hypothesize that the inner radius of the inner disk is at the sublimation radius. The smaller, and hence warmer, silicate dust grains reach the sublimation temperature and only silicate grains larger than several $\mu$m may survive to these temperatures. Small, (sub-)\,$\mu$m-sized carbon dust grains can survive at higher temperatures, and are required in our best-fit model to match the near-IR excess at a distance of about 0.1\,AU, a distance constrained by the interferometric observations. Overall, given the number of free parameters of the model, and that inferring the dust composition on a non-detection of the silicate emission feature is not ideal, we could hardly investigate the dust properties deeper. We assumed all grains to be non-porous, hard spheres, while porosity and possible inclusions of different materials are known to have an impact on the emission features in the mid-IR. 

To reproduce the data, we need the height of the inner wall ($H/r = 0.2$) to be much larger than what hydrostatic equilibrium would prescribe ($H/r \sim 0.02$). This echoes the results of \citet{Thi2011}. The authors used the thermo-chemical code \textsc{Prodimo} to study the morphology of the innermost regions of disk around more massive Herbig Ae stars. They investigated the influence of decoupled gas and dust temperatures. They found the scale height of the innermost regions of the disks to increase significantly compared to models where the temperatures of gas and dust are equal. When decoupled, the gas temperatures are higher than the temperature of the dust grains resulting in a higher inner rim for the gas. Small dust grains will efficiently be dragged along with the gas, which will also lead to a higher scale height for these grains. This finding supports the observational results of \citet{Acke2009} who modeled the SEDs of Herbig Ae/Be stars and concluded that a disk in hydrodynamical equilibrium cannot reproduce the observed near-IR emission. We suggest that it is also true in the case of T\,Cha where the inner rim is more puffed up. The spatial distribution of the smaller grains, which are well coupled to the gas, may increase in the vertical direction and hence provide a larger geometric cross-section to intercept the stellar radiation.

As discussed in \citet[][and references therein]{Schisano2009}, T\,Cha displays a strong variability up to 3\,mag in the $V$-band. The authors discussed the possibility of dusty clumps in the immediate vicinity of the central star, that could partially occult it and be responsible for the observed variability. Such fast variability has also been commonly observed around other young stars (e.g., AA\,Tau, \citealp{Bouvier2007}). Based on CoRot photometric observations in NGC\,2264, \cite{Alencar2010} found that a high fraction of the 83 TTauri stars in their sample (30 to 40\% of stars having inner dusty disks) exhibit photometric time variability comparable to the AA\,Tau phenomenon. The authors suggested this variability is the consequence of a warped inner disk periodically occulting the photosphere. This non-axisymmetric inner disk may trace the magnetospheric interactions between the disk and the star. To account for such variability, \cite{Alencar2010} suggested the scale height at the inner rim to be of the order of $H/r \sim 0.3$ (assuming a random distribution of inclinations). Our finding of a puffed-up inner rim in the innermost regions of T\,Cha, combined with time variability may trace a non axisymmetric, warped inner disk interacting with the stellar magnetosphere.

\section{Modeling the outer disk\label{sec:outer}}

\subsection{Disk parameters\label{sec:disk_outer}}

As mentioned in Sect.\,\ref{sec:geom}, the \textsc{Pionier} $V^2$ present a ``plateau" at low spatial frequencies, that can be explained by an over-resolved, extended contribution. Assuming a 2D Gaussian profile for the photometric field of view of the AT (with a FWHM of about 250\,mas, $d_{\star} = 100$\,pc) material out to 12.5 AU in radius can be detected (depending on the inclination $i$). Therefore the outer disk is in the field of view of the AT and may contribute to the drop of $V^2$ at low spatial frequencies. The same is true for the \textsc{Midi} data where the outer disk acts as a source of incoherent flux and contributes to the drop in correlated flux ($\mathrm{FoV}$ of about 300\,mas). With the setup for the inner disk as described above, we investigated the geometry of the outer disk. Based on all available data and the modeling results presented in \citet{Cieza2011}, we searched for the best model for the outer disk within a grid of models. To limit the number of free parameters, we fixed some of the parameters already explored in \citet{Cieza2011}, namely, the flaring index $\beta = 1.1$, $\alpha=-1$, and $r_\mathrm{out} = 25$\,AU, leaving the following free parameters: the inner radius, the scale height, and the dust mass $M_\mathrm{dust}$. As for the inner disk, several inclinations and position angles for the outer disk are tested against the observations. One should note that the inner and outer disk are assumed to be coplanar (same $i$ and $PA$) in the analysis. The range of values explored in the grid are reported in Table\,\ref{tab:inner}. For the dust, we used the optical constants of \citet{Draine1984}, with $s_{\mathrm{min}}$=0.1\,$\mu$m and $s_{\mathrm{max}}$=3\,mm, assuming $p$=$-3.5$. We did not include amorphous carbon dust grains in the dust content. As opposed to the inner disk, where the lack of emission feature at 10\,$\mu$m required to consider an additional dust population, for the outer disk, there is no observational evidence to support such a choice.

The best model for the outer disk is found by minimizing the un-weigthed reduced $\chi^2_\mathrm{r}$ for the SED (from optical to mm wavelengths), the \textsc{Pionier} $V^2$, and the \textsc{Sam} $V^2$ and closure phases (see Sect.\,\ref{sec:sam}, in which we discuss the best-fit values for $i$ and $PA$). Because the inner disk is slightly different than the one used in \citet[][especially the scale height]{Cieza2011}, we found a lower value for the inner radius of the outer disk ($12 \pm 2$\,AU) to match the increase in the SED long-wards to 15\,$\mu$m. The scale height $H_0$ at a reference radius of $r_0 = 25$\,AU is found to be of about $2.1 \pm 0.2$\,AU, a value close to the one found in \citet{Cieza2011}. We obtained a total dust mass of $8 \times 10^{-5}\,M_{\odot}$ ($5 \times 10^{-5} \leq M_\mathrm{dust} \leq 2 \times 10^{-4}\,M_{\odot}$). With the outer disk being in the $\mathrm{FoV}$ of \textsc{Pionier} at the AT, it introduces a drop in $V^2$ of about 5\% for the lowest spatial frequency in our dataset. Not including the outer disk returns $V^2$ almost equal to unity at these low spatial frequencies, which would be inconsistent with our observations.

\subsection{The outer disk seen in scattered light\label{sec:sam}}

\begin{figure}
\begin{center}
\hspace*{-0.cm}\includegraphics[angle=0,width=1\columnwidth,origin=bl]{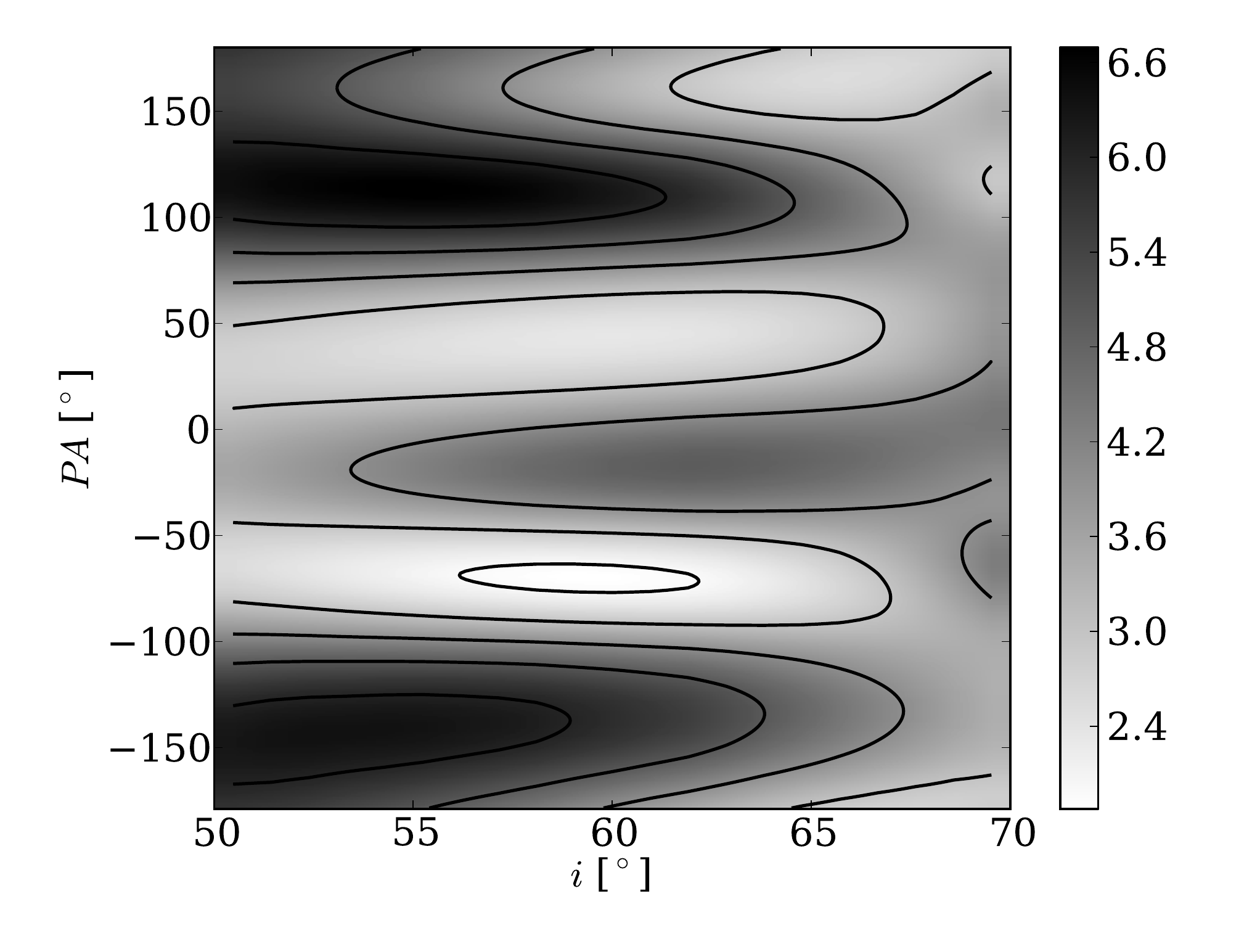}
\caption{\label{fig:PA}Reduced $\chi^2_{\mathrm{r}}$ map and contours, from fitting the \textsc{Sam} closure phases, as a function of the inclination $i$ and position angle $PA$.}
\end{center}
\end{figure}

\begin{figure*}
\begin{center}
\hspace*{-0.cm}\includegraphics[angle=0,width=1.9\columnwidth,origin=bl]{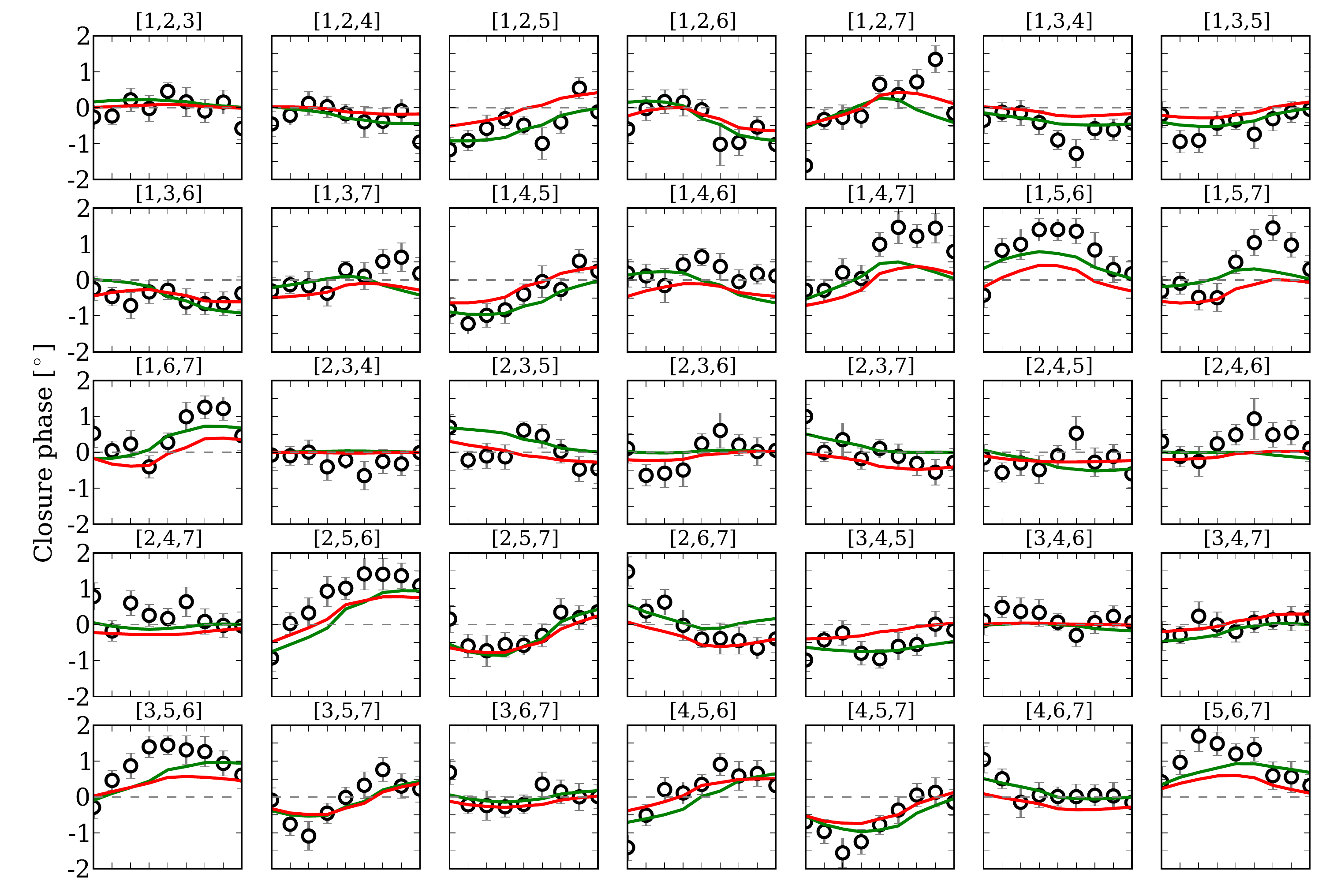}
\caption{\label{fig:TChab}Observed \textsc{Sam} closure phases in black open circles, modeled closure phases using a binary model (green line, solution found by \citealp{Huelamo2011}) and modeled closure phases from our best-fit disk model (red line, $PA = -70^{\circ}$). Above each panels, the corresponding holes triplet of the 7 holes mask is reported.}
\end{center}
\end{figure*}

\begin{figure*}
\begin{center}
\hspace*{-0.cm}\includegraphics[angle=0,width=.6\columnwidth,origin=bl]{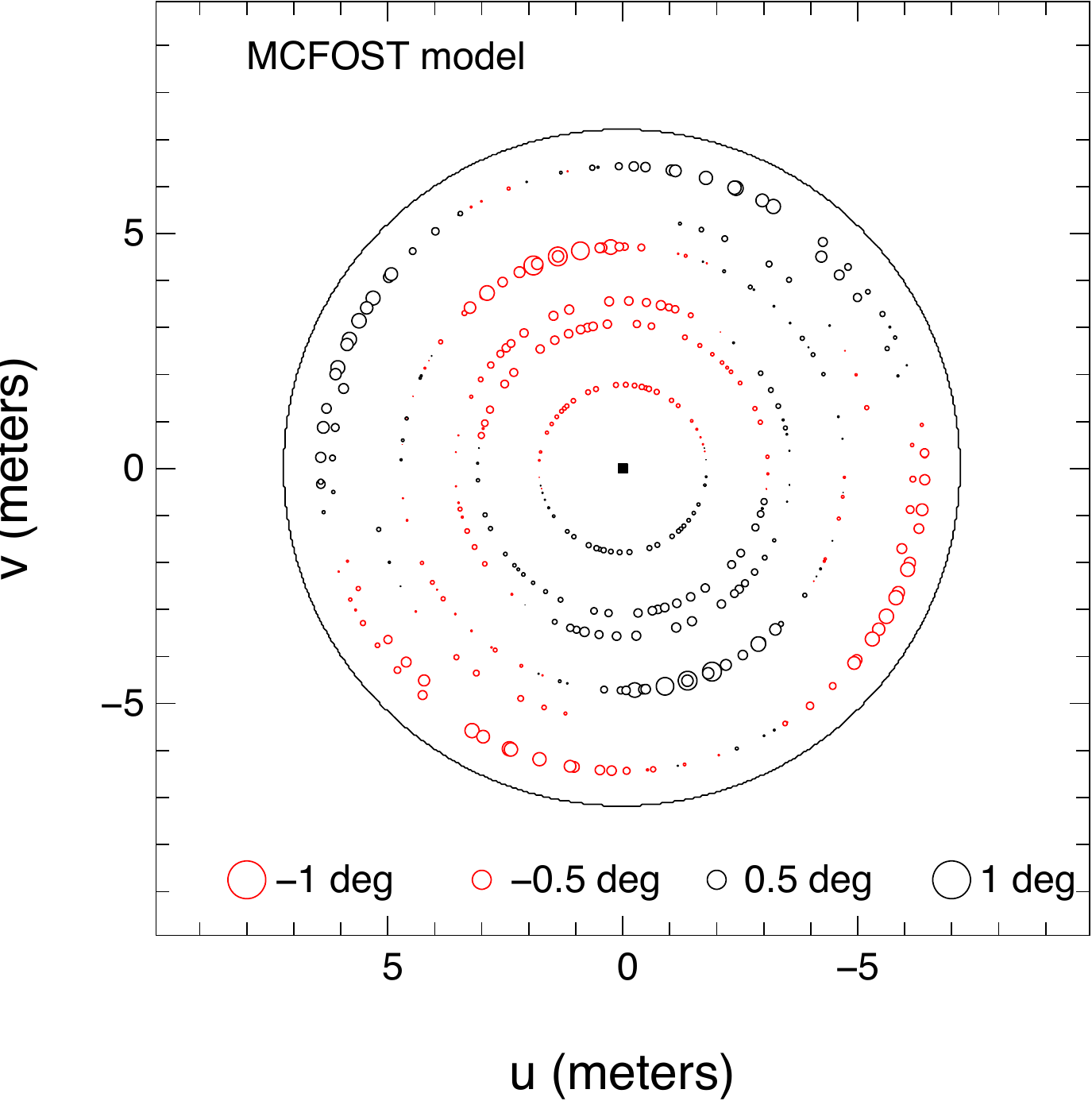}
\hspace*{-0.cm}\includegraphics[angle=0,width=.6\columnwidth,origin=bl]{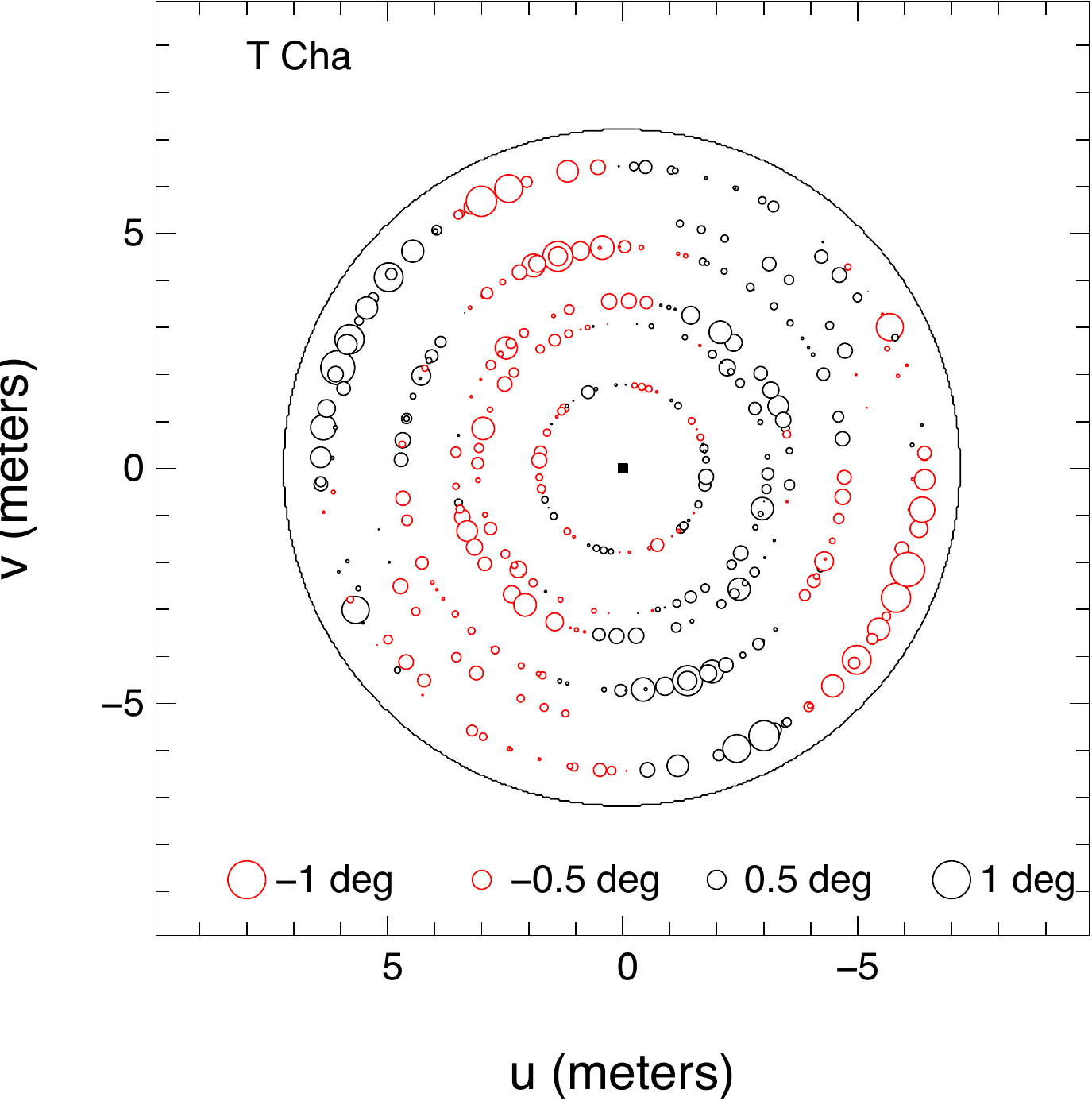}
\hspace*{-0.cm}\includegraphics[angle=0,width=.6\columnwidth,origin=bl]{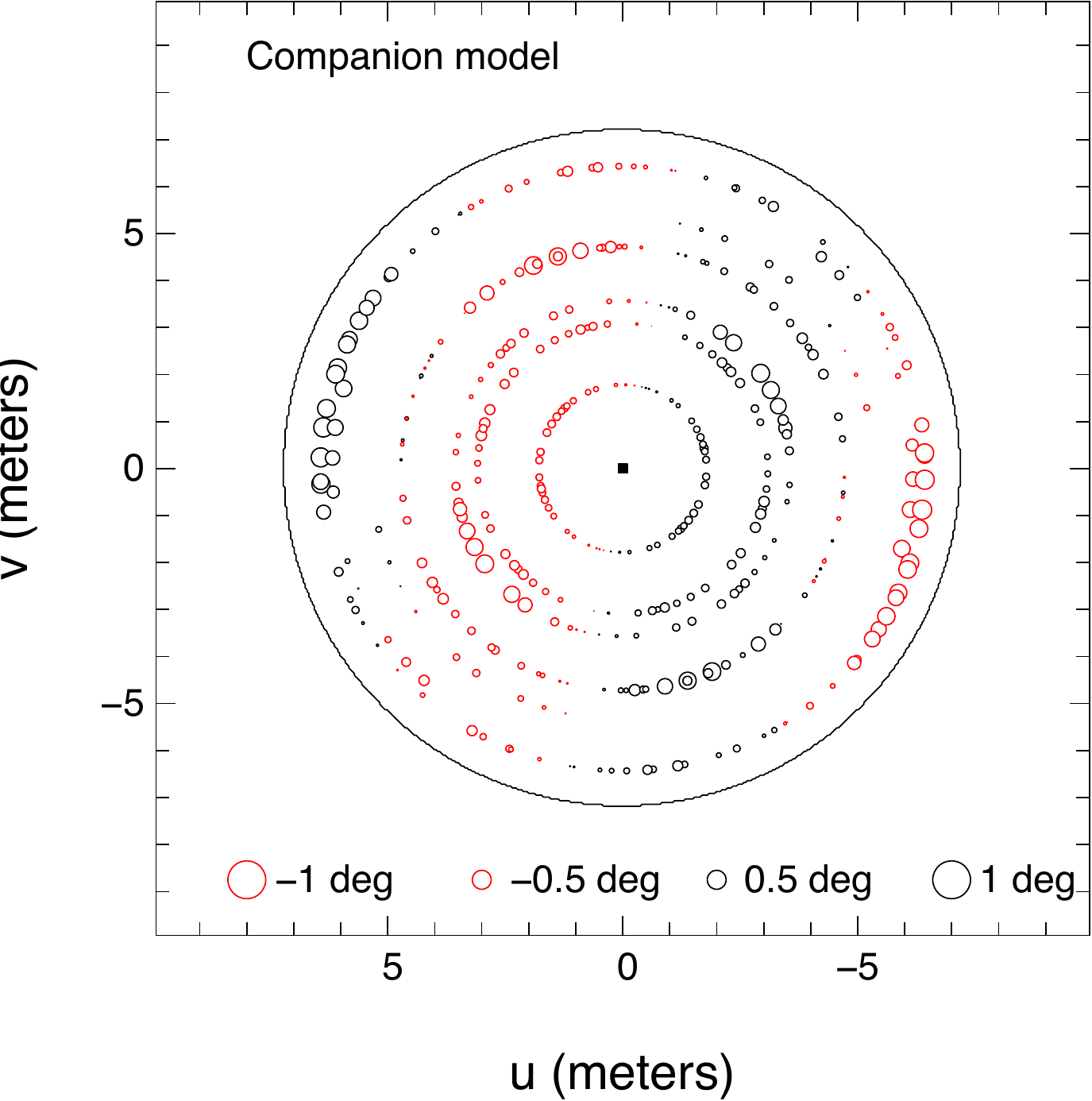}
\caption{\label{fig:phases}Phases representation in the $(u,v)$ plane, for the best-fit \textsc{Mcfost} model (left panel), the original \textsc{Sam} data (middle) and the binary model from \citet[][right]{Huelamo2011}. The color of the circles give the sign of the phases, while their sizes depend on their absolute values.}
\end{center}
\end{figure*}

When observing with \textsc{Sam} the $\mathrm{FoV}$ is of about 500\,mas (50\,AU in diameter at 100\,pc). Previous studies of T\,Cha, and results from the geometric model suggest that the disk is seen at a relatively high inclination ($i \sim 55^{\circ}$, see Sect.\,\ref{sec:inner_mod}). Therefore the wall  of the outer disk ($r_{\mathrm{out}} =12$\,AU) is clearly visible in raytraced images, as well as a large portion of the disk. Because closure phases trace departures from centro-symmetry of the surface brightness, we investigated the closure phase signal generated by the disk alone. We computed $L'$-band closure phases for baseline triplets of the \textsc{Sam} observations, for different inclinations ($50^{\circ} < i < 70^{\circ}$) and position angles ($-180^{\circ} < PA < 180^{\circ}$). One should note that for $PA = 0^{\circ}$, the Eastern part of the disk is inclined towards the observer. Figure\,\ref{fig:PA} shows the reduced $\chi^2_{\mathrm{r}}$ map as a function of both parameters. The minimum reduced $\chi^2_{\mathrm{r}}$ for the \textsc{Sam} data is 1.9, found for $i = 58^{+6}_{-3}$\,$\degr$ and $PA = -70^{\circ} \pm 20\degr$. One should be aware that the \textsc{Pionier} data provided a measurement for the $PA$ of the inner disk with a $180\degr$ degeneracy ($PA = 120\degr$ or $-60\degr$). As a comparison, the reduced $\chi^2_{\mathrm{r}}$ for a single star (i.e., no closure phase signal) is 3.2. From Fig.\,\ref{fig:PA}, one should note that for approximately the same inclination, a local minimum in the $\chi^2_\mathrm{r}$ map is seen for a position angle of about $PA = 40\degr$ ($\chi^2_\mathrm{r} \sim 2.3$). Even the good $(u,v)$ coverage of the \textsc{Sam} observations (see Fig.\,\ref{fig:phases} and its description later in text) is not sufficient to unambiguously distinguish between these two solutions, underlining the need for direct imaging observations. Figure\,\ref{fig:TChab} shows the observed and modeled closure phases with the binary solution found by \citet[][reduced $\chi^2_{\mathrm{r}}$ of 1.5]{Huelamo2011} and our best-fit disk model. Each panel shows consecutive closure phase measurements (observations were repeated 9 times) for a given holes triplet. Given the larger $\mathrm{FoV}$ and lower spatial frequencies of the \textsc{Sam} observations compared to the \textsc{Pionier} instrument, these observations are best suited to constrain the parameters of the outer disk. Consequently, we adopt the aforementioned values for $i$ and $PA$, assuming that the inner and outer disk are coplanar, a reasonable assumption given the uncertainties we find.

To have a better understanding of the origin of the closure phase signal, we computed raytraced images in which the \textsc{Mcfost} code separates different contributions; thermal emission, stellar scattered light, and thermal scattered light. The thermal emission from the outer disk is not responsible for the measured closure phase signal. The outer disk emits negligible amounts at $L'$-band, as it is too cold. However, scattering by the outer disk surface of $L'$-band photons originating from the photosphere and the hotter inner disk produces a significant closure phase signal. These observations probe forward scattering, and therefore constrain the location on the sky of the front side of the disk, that we find to be in the North-East corner. One should note that the phase function of scattering depends, among other parameters, on the grain size. Until now, we used $s_{\mathrm{min}} = 0.1$\,$\mu$m for the silicate dust grains, which is small compared to the wavelength of observations (3.8\,$\mu$m). To test the robustness of our predictions, we increased $s_{\mathrm{min}}$ to 1\,$\mu$m in the outer disk, so that the grains no longer are in the Rayleigh limit. With $2 \pi s > \lambda$, scattering becomes even more anisotropic (e.g., Fig.\,3 of \citealp{Mulders2013}). With this model, the quality of the fit to the closure phases is still in good agreement with the observations ($\chi^2_{\mathrm{r}} = 1.9$ for a $PA$ of $-69^{\circ}$). This suggests that a good fit to the closure phases is achieved as long as light is preferentially scattered in the forward direction. Within the grid of models for the outer disk, we could find slightly better fits to the \textsc{Sam} closure phases ($\chi_\mathrm{r}^2 \sim 1.8$) when increasing the scale height of the outer disk from $H_0 = 2.1$ to 2.3 at $r_0 = 25$\,AU. However, such models overpedict the flux at about 30\,$\mu$m by a factor $\lesssim 2$. Under the assumption the \textsc{Sam} observations trace the outer disk, it could be indicative of dust settling in the outer disk: if $\mu$m-sized grains are present in the uppermost layers of the disk, without larger grains, the SED would remain comparable but anistropic scattering would also happen at higher altitudes. Such models with larger scale height additionally increase the over-resolved contribution of the outer disk in the \textsc{Pionier} $\mathrm{FoV}$, decreasing by a few percentage points the $V^2$ at short baselines. However, without further constraints on the outer disk, investigating dust settling is out of the scope of this paper.

Bottom three panels of Figure\,\ref{fig:phases} show the phases (not closure phases) for the best-fit disk model, the original data, and the binary solution found by \citet[][left to right panels, respectively]{Huelamo2011}. The phases plotted in these three figures are obtained via a linear combination of the closure phases. The relation is obtained by inversion of the closure phase to phase matrix (composed of $1$, $0$ and $-1$). The unknown parameters of the phases (e.g., piston, tip-tilt) are set at zero. The fit to the data is always performed on the closure phases, but the representation of the phases helps to better visualize and compare the observations to the models. The best $PA$ found for the disk and the binary models are quite different ($-70\degr$ vs. $+78\degr$, respectively). Intuitively, one may expect that the semi-minor axis of disk (along which the asymmetry is generated) should be aligned with the $PA$ of the binary, and hence would expect a $\pm 90^{\circ}$ difference between the two best-fit position angles (the $PA$ of the disk tracing its semi-major axis along the North-South direction). In principle, the 7\,holes mask of the \textsc{Sam} instrument can probe 21 different phases. But what is actually measured by the instrument consists of 35 closure phases (15 independent closure phases). Closure phases do not directly trace the phases, and we conclude that closure phases can be reproduced at the same level with different models. The differences in $PA$ is not an inconsistency between the two models.

In the detection paper, the candidate companion was not detected in the $K_{\mathrm{s}}$-band. To check if our model is consistent with this non-detection, we computed closure phases in the $K_{\mathrm{s}}$-band with the same baseline triplets as the $L'$-band data and the same model ($s_{\mathrm{min}} =0.1$\,$\mu$m, $PA = -70^{\circ}$) and we found smaller closure phases compared to the $L'$-band. Overall, we find that only a couple of baseline triplets (13 out of 35) show an absolute closure signal larger than 0.5$^{\circ}$ (and only one triplet with an absolute closure phase greater than 1$^{\circ}$). Depending on the observing conditions and baseline triplets (i.e., projected baseline, which depends on the time of the observations), our model is consistent with a non-detection.

\subsection{The contribution of PAHs\label{sec:PAHs}}

The only emission feature detected in the \textsc{Irs} spectrum of T\,Cha is associated with PAHs, at 11.3\,$\mu$m. In an attempt to obtain the most complete description of the entire system, we included the contribution of PAHs in \textsc{Mcfost}. We used optical constants from \cite{Li2001}, and considered the PAHs to have a single size of $3.55 \times 10^{-4}$\,$\mu$m. Because the  emission of the PAHs is non-thermal, they do not necessarily need to be close to the star to contribute at 11.3\,$\mu$m, and they were therefore included in the outer disk. This choice is supported by the fact that the 11.3\,$\mu$m emission feature is not observed in the \textsc{Midi} correlated fluxes, as opposed to the \textsc{Irs} spectrum (Fig.\,\ref{fig:data}). However, the star must contribute in the UV regime to excite the electronic states of the PAHs. We parametrized the UV flux of the central star via its UV excess ($L_{\mathrm{UV}} / L_{\star} = 0.01$, assuming a spectral shape $L_{\lambda} \propto \lambda ^{0.5}$, see \citealp{Woitke2010}). To reproduce the flux level of the observed 11.3\,$\mu$m feature, we found a mass fraction of 0.5\% of the total outer disk dust mass to be enough. Including the PAHs additionally has a small, but still noticeable impact on the \textsc{Pionier} $V^2$ at low spatial frequencies. As a source of non-thermal emission, PAHs rendered the outer disk slightly brighter even in the $H$-band. Including them in the model produces slightly lower $V^2$ for $B / \lambda < 20 \times 10^{-6}$\,rad$^{-1}$.

However, the best-fit to the \textsc{Sam} closure phases is considerably degraded (reduced $\chi^2_{\mathrm{r}} = 7.3$ compared to 1.9) when including PAHs. We found the best $PA$ to be $118^{\circ}$. The source of asymmetry is this time backward scattering, and the back and front sides of the disk are reversed compared to the previous solution ($PA = -70^{\circ}$). This finds an explanation in the very small size ($s = 3.55 \times 10^{-4}$\,$\mu$m) of the PAHs grains that are included in the best-fit model. Such small dust grains scatter almost isotropically and therefore the wall of the outer disk becomes a significant source of asymmetric emission. Therefore, the surface brightness distribution of the asymmetry becomes clearly different (wall vs. forward scattering in the upper layers of the disk), which explains the difference in $\chi^2_{\mathrm{r}}$. This suggests that the small PAHs grains are not located at the inner edge of the outer disk (see \citealp{Verhoeff2010} for a similar discussion applied to the case of HD\,95881). Because we do not know the exact size and spatial distribution of the PAHs in the outer disk (to be further addressed in Meeus et al. in preparation), we opt not to include them in our best-fit model.

\subsection{Discussion}

The assumption that the \textsc{Sam} observations do, in fact, trace the disk around T\,Cha and not a candidate companion, has some direct implications on the location of the PAHs. These  grains cannot be located at the inner edge of the outer disk, otherwise backward scattering is too efficient. This implies that PAHs should be located further out in the disk, and we hypothesize they have been transported outwards, in the outer disk's upper layers by radiation pressure or stellar winds. 

The mass of the candidate companion is not well constrained and an upper limit of 80\,M$_{\mathrm{Jup}}$ was found by \citet{Huelamo2011}, suggesting a relatively high mass companion. Simulations of a gap opened by a planet suggests the outer disk can become eccentric depending on the planet mass. With a mass above $\sim$\,3\,M$_{\mathrm{Jup}}$, the companion should rapidly push away material that was at the loci of the Lindblad resonances (e.g., 2:1). These resonances are known to efficiently damp any eccentricity and circularize the companion's orbit. If most of the gas has been evacuated from these regions, the damping effect becomes negligible. Additionally, other resonances, such as the 3:1 (if not cleared), will on the contrary excite the companion's eccentricity and the outer disk will therefore quickly lose its axisymmetry. If the outer disk is warped, shows spiral structures (e.g., HD\,135344\,B, \citealp{Muto2012}), or prominent asymmetries in a preferred direction (e.g., HD\,142527, \citealp{Rameau2012}; \citealp{Casassus2012}), such asymmetries may also be responsible for a significant signal in the closure phases.

However, the phase function of scattering is a challenging question. As already mentioned, we used spherical, non-porous grains which may not lead to a proper description of the scattered light contribution. An accurate description (and subsequent modeling) of anisotropic scattered light, asymmetries or clumps in the outer disk requires at least spatially resolved images of the outer disk at different wavelengths (e.g., in the near-IR with NaCo in ADI mode, and in the sub-mm with \textsc{Alma}). 

Concerning the size of the outer disk, as in \citet{Cieza2011}, assuming a surface density with an exponent $\alpha = -1$ we find that the solution of a narrow outer disk ($r_{\mathrm{out}} = 25$\,AU) provides a good match to the SED, especially at Herschel/\textsc{Pacs} and \textsc{Spire} wavelengths. We could find comparable solutions with greater values for the outer radius (300\,AU) but the surface density then has to be extremely steep ($\alpha = -$3) compared to the minimum mass solar nebula and other observed disks, otherwise the Herschel/\textsc{Pacs} and \textsc{Spire} observations would be over-predicted. Because no stellar companion was detected farther out than 20\,AU (very deep imaging of T\,Cha by \citealp{Vicente2011}), this result of a narrow disk cannot be explained by an outward truncation of the outer disk. 

\cite{Pinilla2012} studied the dust distribution in disks sculpted by planets. Usually, mm-sized particles are subjected to a rapid inward drift. However, when a gap-opening planet is included in their simulations, a pressure bump is created at the outer edge of the gap. Such a pressure bump will stop the dust particles as they start experiencing a positive pressure gradient. These mm-sized dust particles will therefore be trapped around the pressure maximum. As a consequence, a significant peak in the density profile will appear at the edge of gap, where grains start to pile up because of the planet-disk interaction. The high concentration of mm-sized grains at the edge of the disk will then dominate the emission in the far-IR, mm-sized grains being efficient emitters at these wavelengths. The outermost regions of the disk (beyond the pressure maximum) will then mostly contain smaller grains ($\lesssim$\,100\,$\mu$m) that contribute less to the far-IR emission. This could mimic a narrow outer disk, even though it extends farther out with a population of sub-mm dust grains.

According to \citet{Pinilla2012}, the dust grains at the inner edge of the outer disk will eventually collide at high velocities. They will fragment and produce smaller dust grains which will follow the gas inside the gap (efficient coupling), and may refill the inner disk.

\section{Size of the gap: a self-shadowed disk?\label{sec:shadow}}

In the best-fit model presented above, the inner disk is extremely narrow, and the gap between the inner and outer disk is therefore very wide. Assuming a single planet lies in the gap, theories of planet-disk interactions cannot explain our observational result. The size of the gap opened by a single planet, in the gas profile, can firstly be approximated as twice the Hill radius $r_{\mathrm{H}} = r_{\mathrm{p}} (M_{\mathrm{p}}/ 3M_{\star})^{1/3}$. Even with a mass as large as 80\,$M_{\mathrm{Jup}}$, at a projected distance of 6.7\,AU as suggested by \citet{Huelamo2011}, this would lead to a gap size of about 3.5\,AU (\citealp{Zhu2012}). Within the uncertainties on the position angle of the planet relative to the disk and the inclination of the entire system, the inner radius of the outer disk is consistent with this value. However, the outer radius of the inner disk does not fit in this picture. 

We find the inner disk to be optically thick, with an integrated optical depth in the midplane of $\tau = 2$ in the $H$-band. Material can therefore lie in the shadow of the inner disk. To address the question of how much mass can be hidden, we included an additional disk in the \textsc{Mcfost} best model, right behind the inner disk, to mimic an larger, self-shadowed disk. For the dust composition, we also used the optical constants of \citet{Draine1984}, with $s_{\mathrm{min}} = 0.1$\,$\mu$m and $s_{\mathrm{max}} = 1$\,mm, assuming $p = -3.5$. We ran several models where we increased the outer radius between 1 and 5\,AU by steps of 1\,AU. For each $r_{\mathrm{out}}$ value, we increased the mass of dust until the fit becomes inconsistent with the data. It appears that the SED and the \textsc{Midi} data are best suited to constrain the dust mass in the self-shadowed disk. Indeed, the \textsc{Pionier} data are sensitive to the inner radius of the inner disk, while the SED will immediately reveal if there is any emission in excess in the mid-IR. For a given $r_{\mathrm{out}}$, we stopped increasing the dust mass once the modeled correlated fluxes are no longer within the 3$\sigma$ uncertainties of the observed correlated fluxes. We repeated this exercise for two different scale height reference values ($H_0 = 0.05$, and 0.1\,AU at $r_0 = 1$\,AU). Overall, the dust mass that can be hidden in the shadow of the inner disk varies from $\sim 5 \times 10^{-11}$ up to $\sim 2 \times 10^{-9}$\,$M_{\odot}$ for $r_{\mathrm{out}} = 1$ and 5\,AU, respectively (2.5 and 100 times the mass of the inner disk). With the available dataset, we cannot further constrain the extent nor the mass of this self-shadowed disk.

\subsection{The best-fit model}

\begin{figure*}
\begin{center}
\hspace*{-0.cm}\includegraphics[angle=0,width=2\columnwidth,origin=bl]{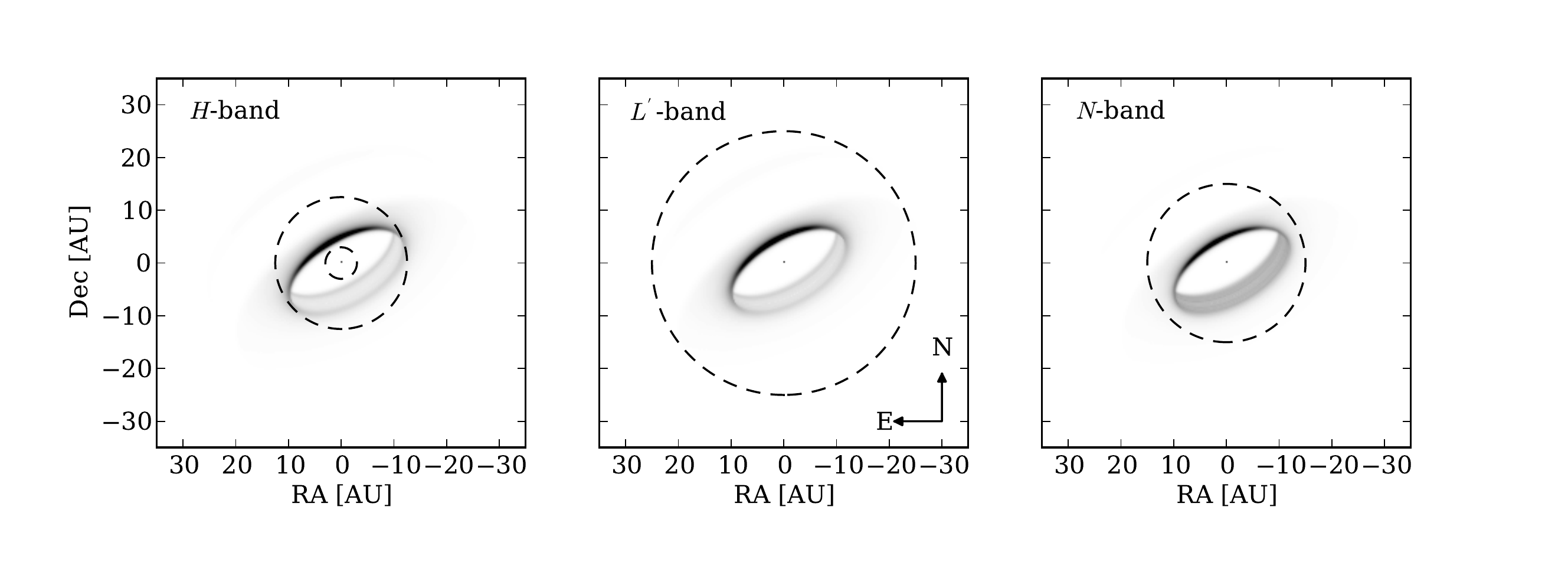}
\caption{\label{fig:image}Color-inverted raytraced images of the best-fit model, in the $H$-, $L'$-, and $N$-band (left to right, respectively). The position angle of the disk is $-70^{\circ}$. The front side of the disk is in the North-East direction, where forward scattering is observed,  while the back side and the wall of the outer disk are in the South-West direction. On each panels the dashed circles represent the $\mathrm{FoV}$ of the different instruments: \textsc{Pionier} and \textsc{Amber} for the $H$-band (large and small circles, respectively), \textsc{Sam} for the $L'$-band, and \textsc{Midi} for the $N$-band.}
\end{center}
\end{figure*}

\begin{figure}
\begin{center}
\hspace*{-0.cm}\includegraphics[angle=0,width=\columnwidth,origin=bl]{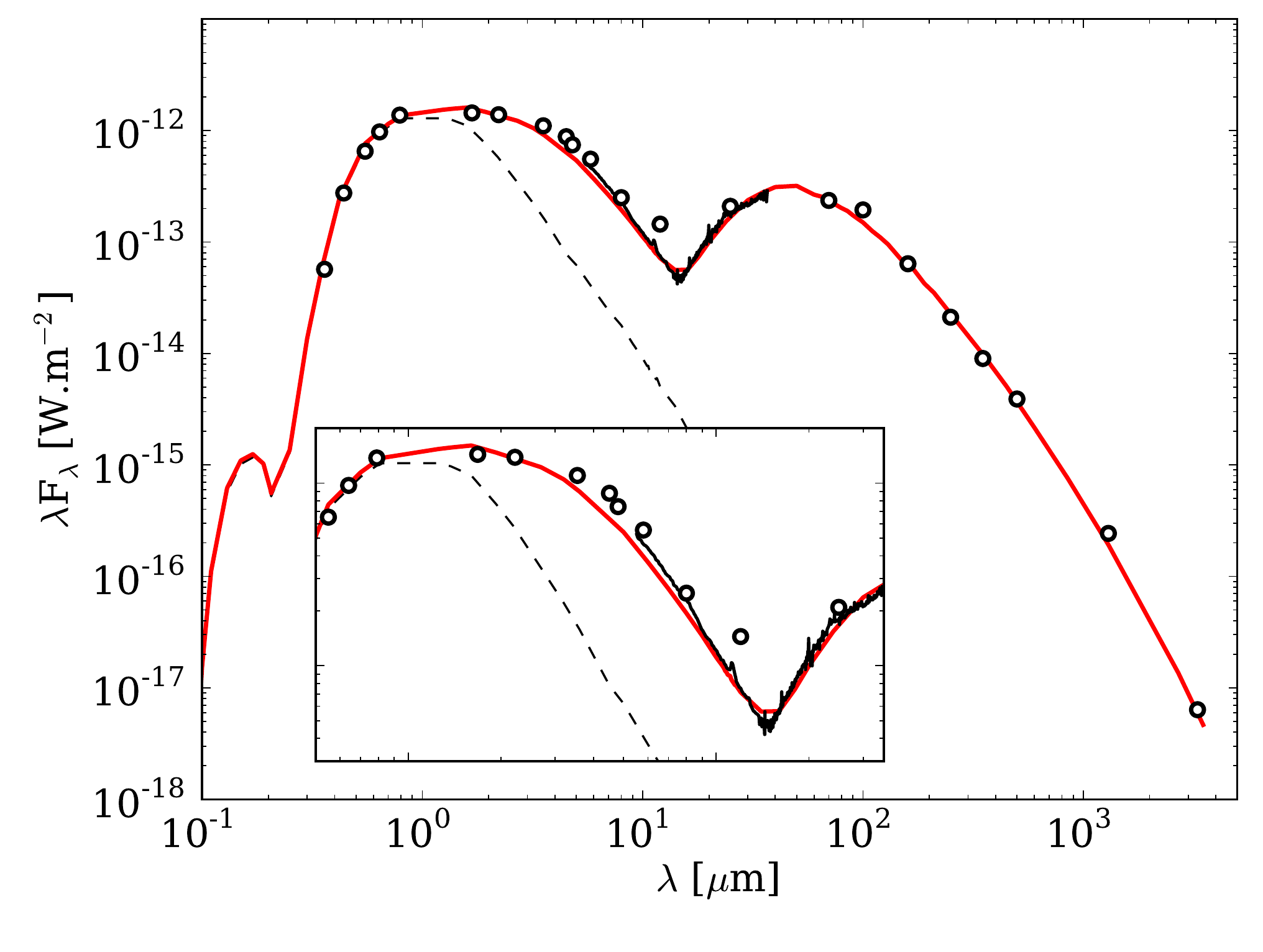}
\caption{\label{fig:SED}Best-fit \textsc{Mcfost} model (in solid red) to the SED of T\,Cha (photometric observations in black, opened circles), including a zoom on the 0.5--35\,$\mu$m spectral range. The dashed line is the photosphere.}
\end{center}
\end{figure}

\begin{table*}
\caption{Parameters for the best-fit \textsc{Mcfost} model. Parameters without uncertainties were fixed in the search for the best-fit model. See text for details.\label{tab:model}}
\begin{center}
\begin{tabular}{lccccccccccc}
\hline \hline
& $r_{\mathrm{in}}$ & $r_{\mathrm{out}}$ & $\alpha$ & $\beta$ & $H_0$  & $r_0$ & $M_{\mathrm{dust}}$ & Dust & Mass frac. & $s_{\mathrm{min}}$ & $s_{\mathrm{max}}$ \\
& [AU] & [AU] & & & [AU]  & [AU] & [$M_{\odot}$] & comp. & [\%] & [$\mu$m] & [$\mu$m] \\
\hline
Inner & $0.07 \pm 0.01$ & $0.11 \pm 0.02$ & -1 & 1 & $0.02 \pm 0.0025$ & 0.1 & $2_{-1}^{+2} \times 10^{-11}$ & Astrosil & 70 & 5 & 1000 \\
 & & & & & & & & Carbon & 30 & 0.01 & 1000\\
Outer & $12 \pm 2$ & 25 & -1 & 1.1 & $2.1 \pm 0.2$ & 25 & $8_{-3}^{+12} \times 10^{-5}$ & Astrosil & 99.5 & 0.1 & 3000 \\
 & & & & & & & & PAHs & 0.5 & $3.55 \times 10^{-4}$ & - \\
 \hline
\end{tabular}
\end{center}
\end{table*}

\begin{figure*}
\begin{center}
\hspace*{-0.cm}\includegraphics[angle=0,width=1\columnwidth,origin=bl]{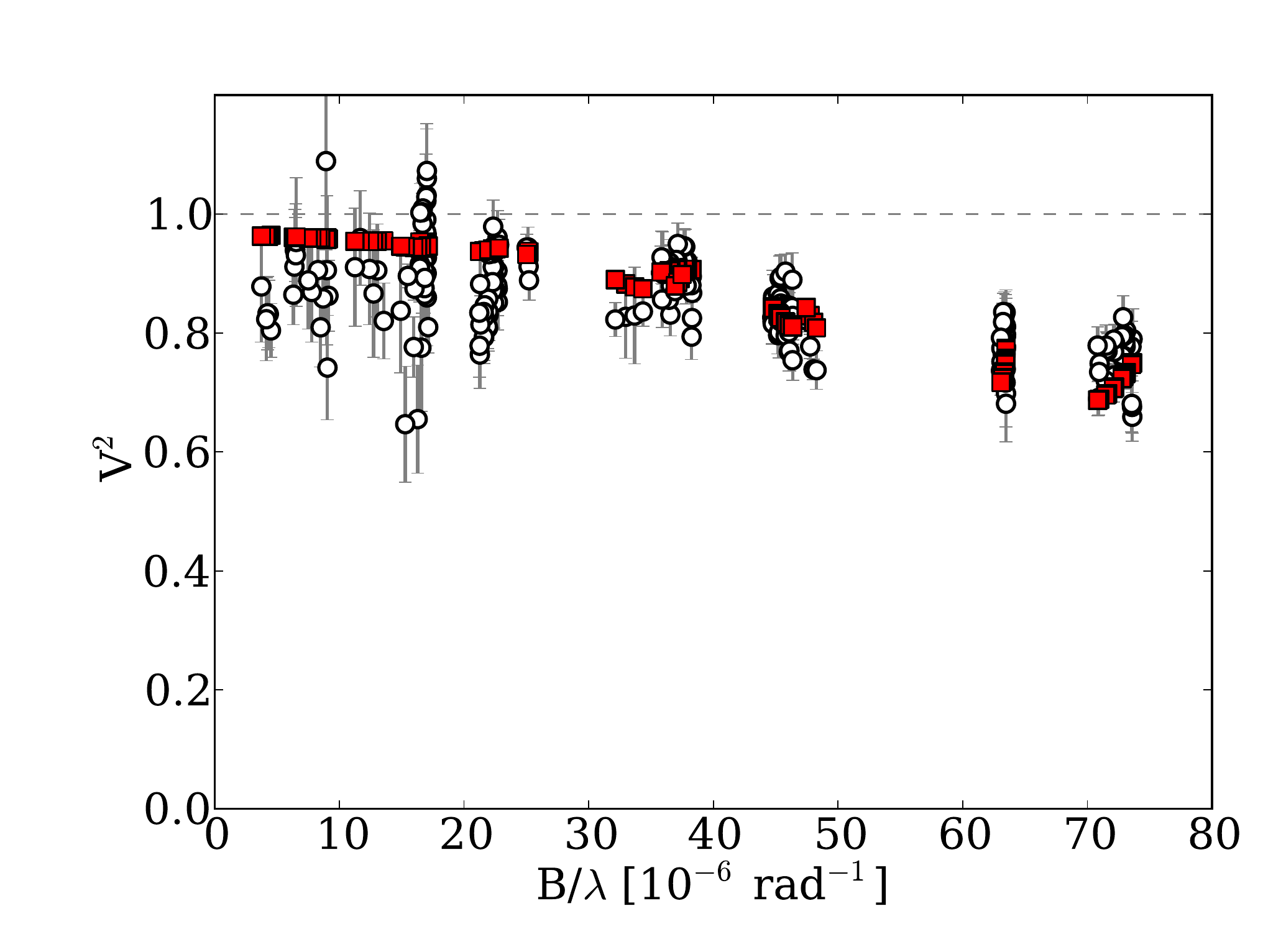}
\hspace*{-0.cm}\includegraphics[angle=0,width=1\columnwidth,origin=bl]{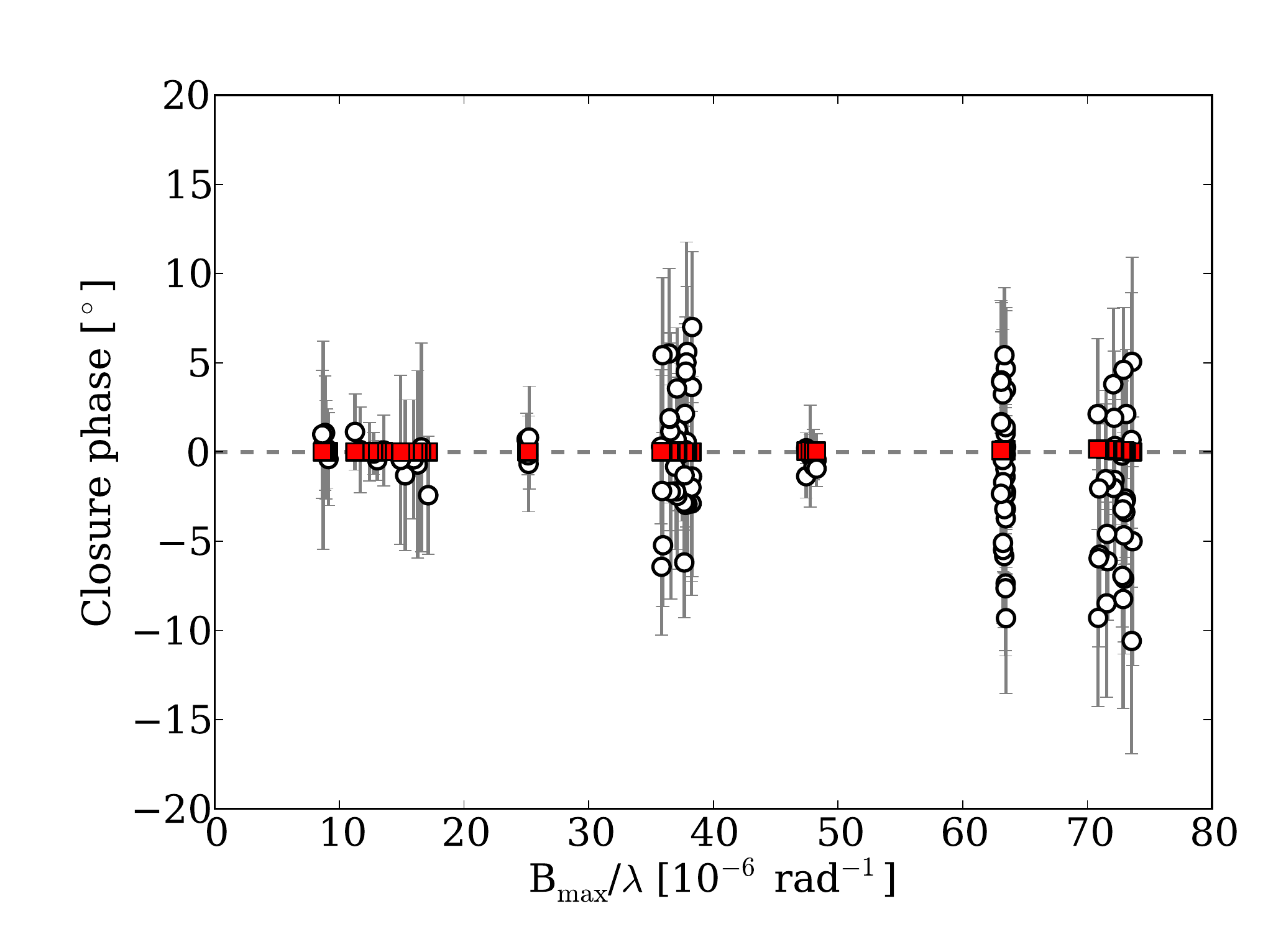}
\hspace*{-0.cm}\includegraphics[angle=0,width=1\columnwidth,origin=bl]{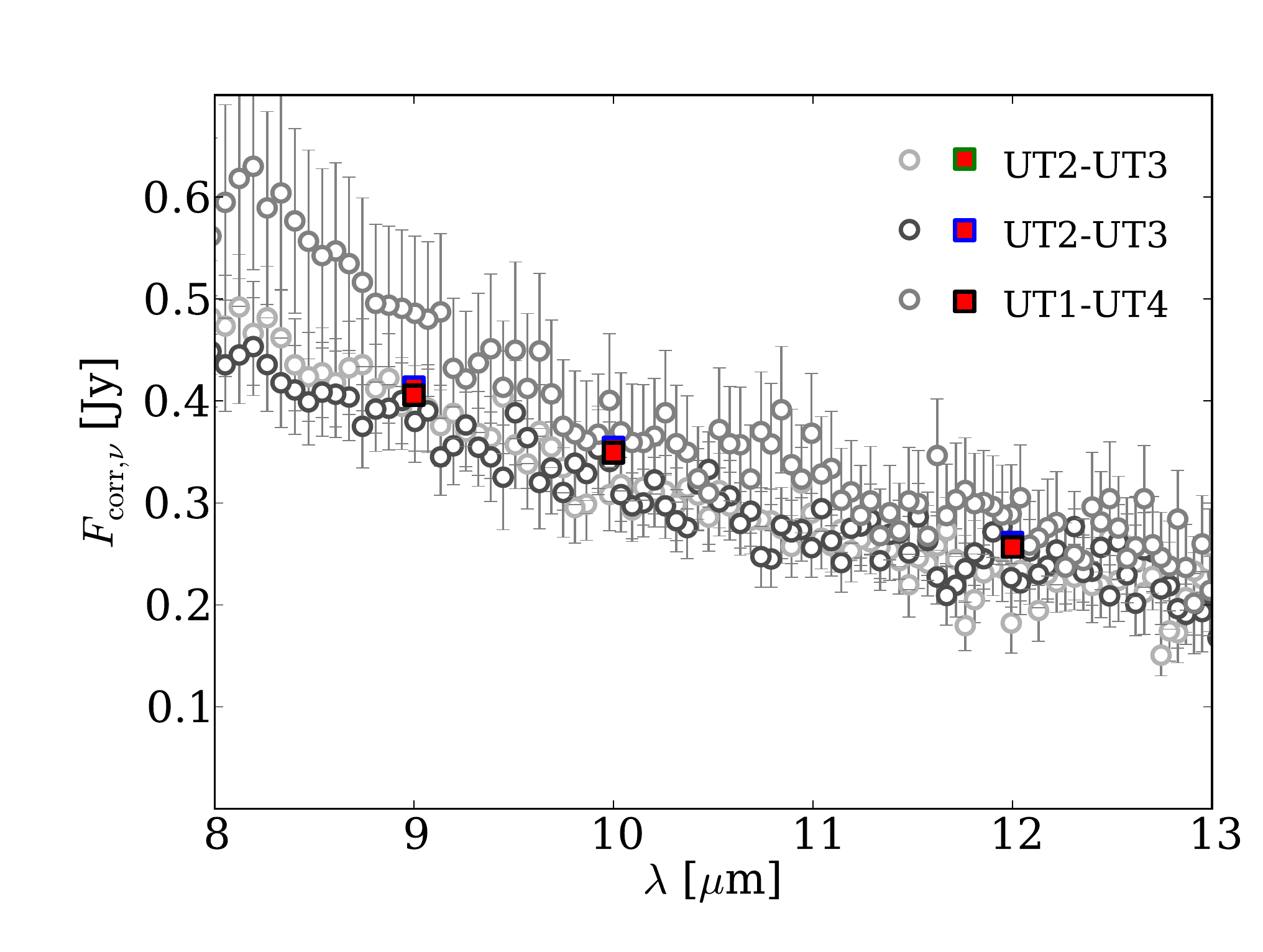}
\hspace*{-0.cm}\includegraphics[angle=0,width=1\columnwidth,origin=bl]{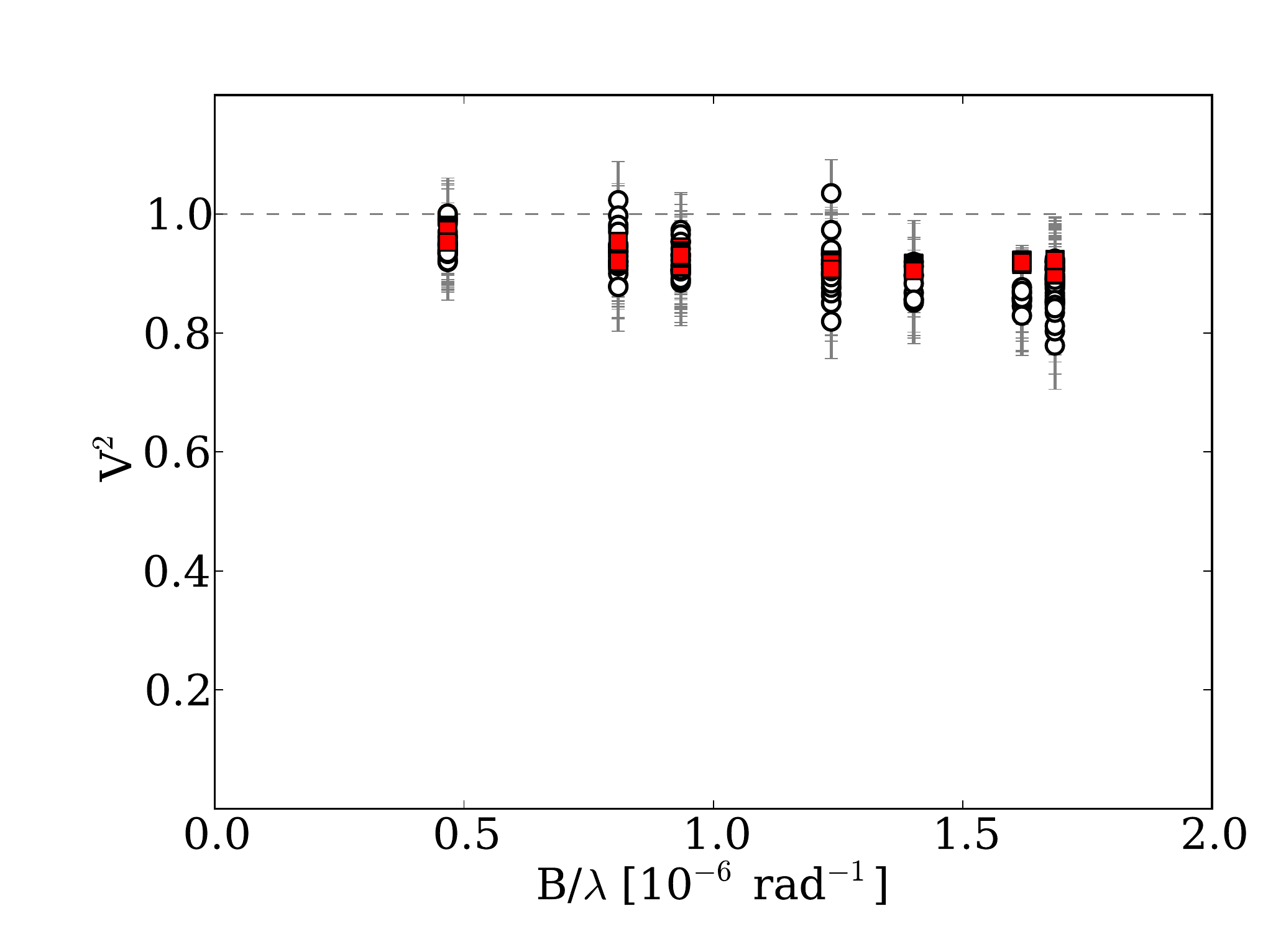}
\caption{\label{fig:mcfost}{\it Upper left:} Observed and modeled \textsc{Pionier} $V^2$ (black circles, red squares, respectively) for the best-fit \textsc{Mcfost} model. {\it Upper right:} same for \textsc{Pionier} closure phases. {\it Bottom left:} observed \textsc{Midi} correlated fluxes (grey circles) and modeled correlated fluxes at 9, 10, and 12\,$\mu$m (red squares). {\it Bottom right:} observed and modeled \textsc{Sam} $V^2$ (black circles, red squares, respectively).}
\end{center}
\end{figure*}

\begin{figure}
\begin{center}
\hspace*{-0.cm}\includegraphics[angle=0,width=\columnwidth,origin=bl]{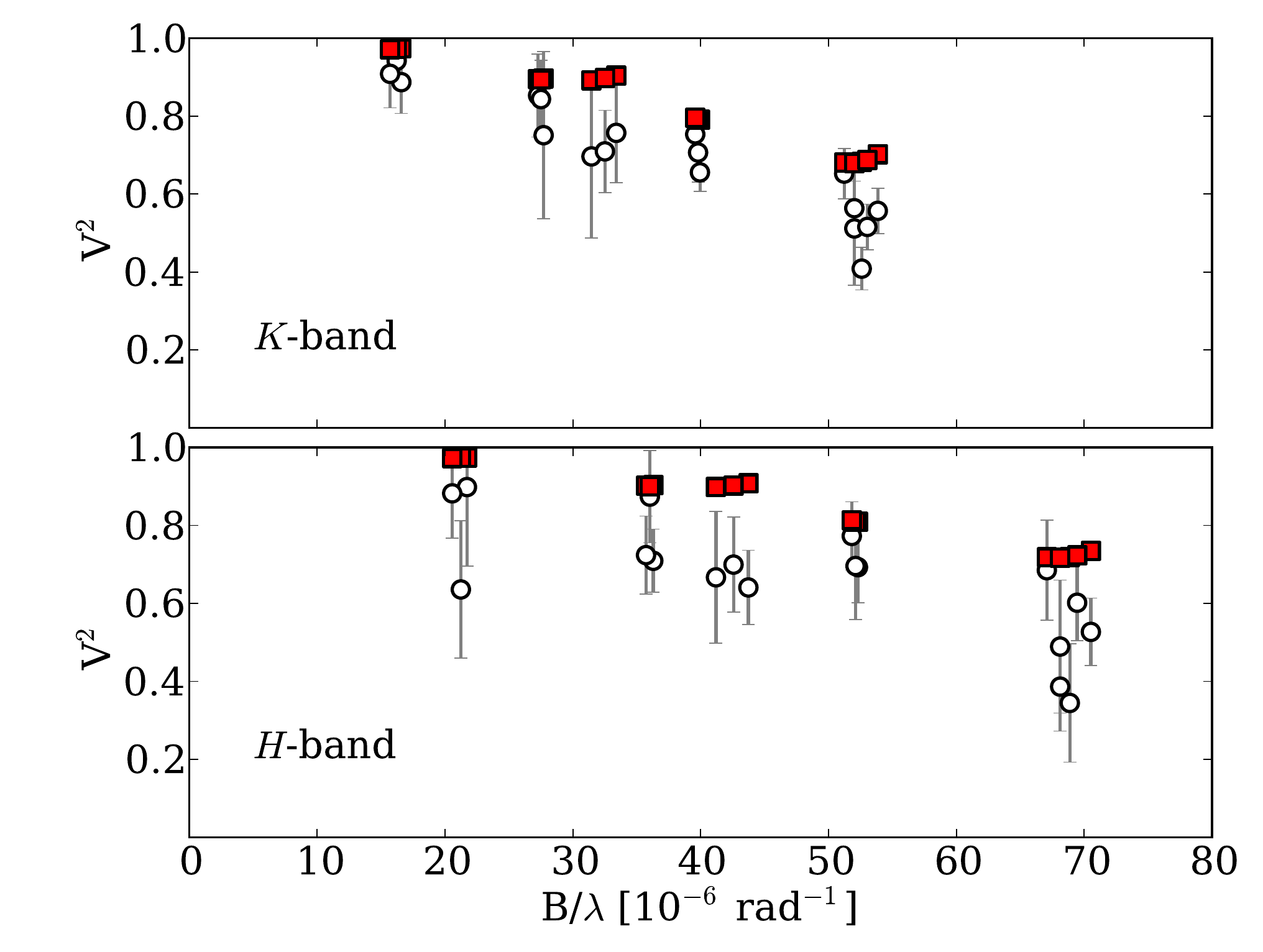}
\caption{\label{fig:amber}Observed and modeled visibilities (open circles and red squares, respectively) for the best fit-model compared to the VLTI/\textsc{Amber} observations presented in \citet{Olofsson2011}. Top and bottom panels display $K$- and $H$-band data, respectively.}
\end{center}
\end{figure}

\begin{figure}
\begin{center}
\hspace*{-0.cm}\includegraphics[angle=0,width=\columnwidth,origin=bl]{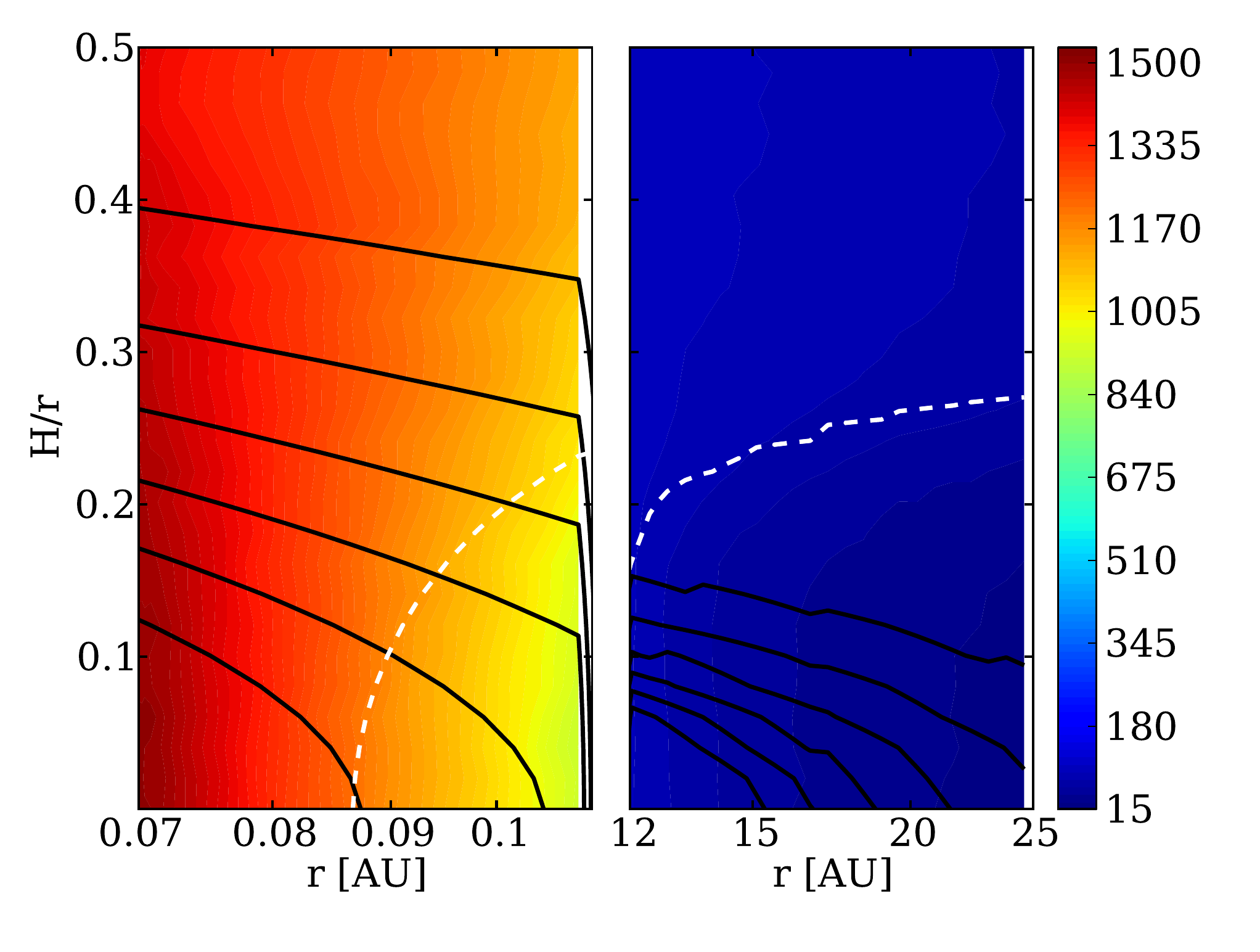}
\caption{\label{fig:maps}Temperature maps for the inner and outer disks (left and right, respectively). The dark contours represent $[$0.6,0.5,0.4,0.3,0.2,0.1$]$ times the maximum density, and the white dashed line represent where the integrated radial optical depth equals 1.}
\end{center}
\end{figure}

Figure\,\ref{fig:image} shows color-inverted raytraced images of the best-fit model in the $H$-, $L'$-, and $N$-band (left to right, respectively), with $PA = -70^{\circ}$, $i = 58^{\circ}$. The $\mathrm{FoV}$ of the different instruments are shown as dashed circles on each panels: \textsc{Pionier} and \textsc{Amber} in the $H$-band (250 and 60\,mas, respectively), \textsc{Sam} in the $L'$-band (500\,mas), and \textsc{Midi} for the $N$-band (300\,mas). A detailed comparison of the best-fit model discussed above with the various data sets is presented on Figs\,\ref{fig:SED} to \ref{fig:amber}. Fig.\,\ref{fig:SED} shows the best-fit model to the SED of T\,Cha from optical to mm wavelengths, including a zoom on the 0.5--35\,$\mu$m spectral region. Upper panels of Figure\,\ref{fig:mcfost} show the best-fit model, assuming a $PA$ of $-70^{\circ}$ and an inclination of $58^{\circ}$, to the \textsc{Pionier} $V^2$ (left) and closure phases (right). Bottom panels of Figure\,\ref{fig:mcfost} shows the corresponding \textsc{Midi} correlated fluxes calculated at 9, 10 and 12\,$\mu$m (left) and the \textsc{Sam} $V^2$ (right). Figure\,\ref{fig:amber} shows the \textsc{Amber} data presented in \citet{Olofsson2011} in $K$- and $H$-band (top and bottom panels, respectively) along with the best fit-model for comparison. Even though the best-fit model is in decent agreement with the data, the modeled visibilities tend to over-predict the observed $V^2$. This may be the consequences of the source variability in the innermost regions (e.g., \citealp{Espaillat2011}). Table\,\ref{tab:model} summarizes the disk parameters and dust composition used in the best-fit model. Finally, Figure\,\ref{fig:maps} shows the temperature map for the inner and outer disks.

\section{Conclusion\label{sec:discuss}}

In this study, we presented a detailed \textsc{Mcfost} model that can account for all available interferometric and photometric observations of the 7\,Myr old transitional disk around T\,Cha. We find the geometry of the circumstellar material to be (at least) twofold. First, a narrow inner disk must be located close to the star and is responsible for the near-IR excess as well as the drop in $V^2$ at high spatial frequencies. Based on the significant near-IR excess, we concluded its scale height to be important, increasing its geometric section, so that a large number of the dust grains have high temperatures ($\sim$\,1500\,K). The lack of solid-state emission features in the mid-IR \textsc{Irs} spectrum of T\,Cha suggests that $\mu$m-sized silicate grains are at their sublimation temperature and are evaporating. Small carbon grains (with a higher sublimation temperature) may survive in these regions of the disk and help reconciling the interferometric observations and the SED. We also discussed the possibility of a more extended (up to several AU) self-shadowed inner disk. 

Second, we find that the outer disk (possibly narrow) starts at around 12\,AU, and is responsible for the far-IR excess. Its presence in the \textsc{Pionier} $\mathrm{FoV}$ slightly decreases the $V^2$ at low spatial frequencies via over-resolved scattered light emission. The small extent of the outer disk, inferred from SED modeling could be the consequence of mm-sized grains accumulating in a pressure maximum at the edge of the gap, caused by a relatively massive companion ($\geq 1$\,M$_{\mathrm{Jup}}$) opening a gap. The outer disk may, however, be more extended in the population of $\lesssim$\,100\,$\mu$m-sized grains.

Finally, our best-fit model indicates that the disk itself is capable of producing a significant closure phase signal, due to forward scattering. With present-day observations, constraints on the outer disk's morphology, and the scattering phase function, we suggest that the \textsc{Sam} closure phase measurements trace the outer disk in scattered light and provide a good constraint on its position angle and inclination. Several key features of the disk around T\,Cha suggest a gap is being opened by a planet, that, however, still needs to be unambiguously detected. More generally, we stress the overall challenge of observing companions in massive bright disks using small-baseline interferometric measurements and model-dependent technics, as asymmetries of the disk (e.g., anisotropic scattering) can strongly contaminate the planet's signatures.

\begin{acknowledgements}
The authors are grateful to the anonymous referee for useful advices and comments that improved the readability of the paper. The authors thank Bertram Bitsch for valuable discussions. C.~P. acknowledges funding from the European Commission's 7$^\mathrm{th}$ Framework Program (contract PERG06-GA-2009-256513) and from Agence Nationale pour la Recherche (ANR) of France under contract ANR-2010-JCJC-0504-01. CNRS is acknowledged for having supported this work in the form of Guaranteed Time Observations (program 089.C-0537(A)).
\end{acknowledgements}

\bibliography{biblio}
\end{document}